\documentclass[fleqn,usenatbib]{mnras}

\usepackage{newtxtext,newtxmath}
\usepackage[T1]{fontenc}
\usepackage{ae,aecompl}

\usepackage{graphicx}	
\usepackage{amsmath}	
\usepackage{cleveref}
\usepackage{subfiles}
\usepackage{subfig}
\usepackage{float}
\usepackage[utf8]{inputenc}		
\usepackage{pifont}
\usepackage{romannum}
\usepackage{array}
\usepackage{placeins}
\usepackage{url}
\usepackage{relsize}
\usepackage{orcidlink}
\usepackage{xcolor}

\newcolumntype{C}{>{\centering\arraybackslash}m{0.12\linewidth}}
\newcolumntype{A}{>{\centering\arraybackslash}m{0.1\linewidth}}
\newcolumntype{L}{>{\centering\arraybackslash}m{0.2\linewidth}}
\newcommand{\ev}{$(\mathlarger{\mathlarger{\mathlarger{\varepsilon}}})$ $ $}
\newcommand{\lev}{${\rm log}$\ev}
\newcommand{\dlev}{$\Delta$\lev}

\title[Substructure or Angular Complexity?]{Galaxy Mass Modelling from Multi-Wavelength JWST Strong Lens Analysis: Dark Matter Substructure, Angular Mass Complexity, or Both?}

\author[Samuel C. Lange et al.]{Samuel C. Lange$^{1}$\thanks{E-mail: samuel.c.lange@durham.ac.uk}\orcidlink{0009-0007-0679-818X}, 
Aristeidis Amvrosiadis$^{1}$,
James W. Nightingale$^{3,1,2}$, 
Qiuhan He$^{1}$,  \newauthor
Carlos S. Frenk$^{1}$,
Andrew Robertson$^{4}$,
Shaun Cole$^{1}$, 
Richard Massey$^{1,2}$, 
Xiaoyue Cao$^{5,6}$, \newauthor
Ran Li$^{5,6}$, 
Kaihao Wang$^{5,6}$
\vspace{4mm}\\
$^{1}$ Institute for Computational Cosmology, Department of Physics, Durham University, South Road, Durham DH1 3LE, UK \\
$^{2}$ Centre for Extragalactic Astronomy, Department of Physics, Durham University, South Road, Durham DH1 3LE, UK \\
$^{3}$School of Mathematics, Statistics and Physics, Newcastle University, Newcastle upon Tyne, NE1 7RU, UK \\
$^{4}$ Observatories, Carnegie Institution for Science, 813 Santa Barbara Street, Pasadena, CA 91101, USA \\
$^{5}$ School of Astronomy and Space Science, University of Chinese Academy of Sciences, Beijing 100049, China \\
$^{6}$ National Astronomical Observatories, Chinese Academy of Sciences, 20A Datun Road, Chaoyang District, Beijing 100012, China
}

\date{Accepted XXX. Received YYY; in original form ZZZ}

\pubyear{2024}

\begin{document}
\label{firstpage}
\pagerange{\pageref{firstpage}--\pageref{lastpage}}
\maketitle 

\pagenumbering{arabic}
\begin{abstract}
We analyze two galaxy-scale strong gravitational lenses, SPT0418-47 and SPT2147-50, using JWST NIRCam imaging across multiple filters. To account for angular complexity in the lens mass distribution, we introduce multipole perturbations with orders $m=1, 3, 4$. Our results show strong evidence for angular mass complexity in SPT2147, with multipole strengths of 0.3-1.7\% for $m=3, 4$ and 2.4-9.5\% for $m=1$, while SPT0418 shows no such preference. We also test lens models that include a dark matter substructure, finding a strong preference for a substructure in SPT2147-50 with a Bayes factor of \dlev $\sim 60$ when multipoles are not included. Including multipoles reduces the Bayes factor to \dlev $\sim 11$, still corresponding to a $5\sigma$ detection of a subhalo with an NFW mass of $\log_{10}(M_{200}/M_{\odot}) = 10.87\substack{+0.53\\ -0.71}$. While SPT2147-50 may represent the fourth detection of a dark matter substructure in a strong lens, further analysis is needed to confirm that the signal is not due to systematics associated with the lens mass model.
\end{abstract}

\begin{keywords}
gravitational lensing: strong -- cosmology: dark matter -- cosmology: observations -- galaxies: structure
\end{keywords}

\section{Introduction}\label{intro}

The nature of the dark matter (DM) has long been one of the most fundamental questions in cosmology. Cold Dark Matter (CDM) is considered the most probable explanation for this "missing matter" (\citealt{DEFW_85}; see \citealt{FW12cdm} for a review). This proposal, together with the assumption of a non-zero Cosmological Constant to explain the accelerated expansion of the Universe, makes up the standard $\Lambda$CDM cosmology that is used to model the Universe \citep[e.g.][]{planck18}. 

Structure formation in a CDM Universe has been intensely studied in cosmological simulations \citep[e.g.][]{Frenk_88,springel2005simulations}. There are still alternatives to CDM that have not been excluded, such as Self-Interacting Dark Matter \citep[SIDM, reviewed in e.g.][]{tulin2018dark} and Warm Dark Matter \citep[WDM, e.g.][]{bode2001halo,Lovell14SHMF}. The formation of low-mass structures provides a key test of CDM and these alternatives; specifically, the CDM paradigm predicts a large population of low-mass structures and of substructures of main halos, that form in a `bottom-up' hierarchical merging process \citep[e.g.][]{FWED1985cdm, diemand2008clumps, springel2008aquarius}. By contrast, in WDM the primordial spectrum of density perturbations has a cutoff on small scales that gives rise in a cutoff in the mass function of halos at a mass corresponding to dwarf galaxies, $\log_{10} ({\rm M/M_\odot}) \sim 8.5$, for the most popular values of the WDM particle mass. This mass function in both CDM and WDM models has been extensively studied - see e.g. \citet{bode2001halo, gao2004shmf,Maccio_10,Lovell14SHMF}, or \citet{zavala2019hmf} for a review. There are several methods for investigating the nature of dark matter. These include measuring the `clumpiness' of the Lyman-$\alpha$ forest \citep{Viel_05,Viel_13}, the abundance of dwarf galaxies, such as the satellites of the Milky-Way \citep[e.g.][]{Newton_21,Newton_24,Enzi_21}. 

Another method is to search directly for low-mass dark halos using gravitational imaging of strong lenses \citep[e.g.][]{vegetti2009potential, vegetti2023strong}. In strong lensing, light rays from a background source are deflected by a massive foreground object, forming multiple images of the source. In the regime of galaxy-galaxy strong lensing, the background source is extended and light rays from different regions of the source are traced through different regions of the lensing galaxy allowing the structure in this lensing galaxy to be studied at high resolution. If a dark substructure happens to lie along one of the lines traced by the source light rays, it will cause a perturbation of the image in an asymmetric manner. All of this can be modelled to create a picture of the lens and source galaxy that best fits the observed image (usually in a Bayesian framework). 

There have been three detections of dark matter structures thus far using this type of gravitational imaging \citep{vegetti2010detection, vegetti2012detection, hezaveh2016detection}, as well as a detection of an Ultramassive Black Hole \citep{nightingale2023smbh}. The inferred halo masses from the three detections depend upon the profile of the dark matter halo assumed in the modelling, but correspond to NFW-halo masses \citep{NFW} in the region $10^9 - 10^{10} {\rm M}_{\odot}$\footnote{Detections from \citet{vegetti2010detection, vegetti2012detection} use tidally truncated pseudo-jaffe profiles, with quoted subhalo masses $\sim 3.5 \times 10^9 {\rm M}_{\odot}$ and $\sim 1.9 \times 10^8 {\rm M}_{\odot}$}. Non-detections of dark matter halos/subhalos (where one would be expected) using gravitational imaging have been used to constrain the subhalo mass function (SHMF) and the possible nature of sterile neutrinos as WDM (e.g. \citealt{ritondale2019nondetect}). This analysis requires a `Sensitivity Map' \citep[e.g.][]{despali2022sensitivity} which indicates where in an image any substructures would be detectable. DM substructures may also be at any redshift along the line-of-sight (rather than at the redshift of the main lens) and can have concentrations that vary from the standard mass-concentration relationship \citep[e.g.][]{li2016constraints, despali2018LOS, minor2021inferring, amorisco2021concentration, he2022LOS, ORiordan2023euclid}. 

Understanding the complexity in the primary lens galaxy is crucial for analyzing dark matter substructures, since overly simplistic mass models can result in false-positive `detections' of substructure. This issue arises due to a degeneracy between the mass complexity of the main deflector and the presence (or absence) of subhalos \citep[e.g.][]{he2023BPL, oriordan2023multipoles}. Multipole perturbations \citep{chu2013multipole} to the standard power-law mass profile \citep{tessore2015EPL} help address some of this complexity. This degeneracy is also encountered when using flux ratio anomalies (brightness differences in multiple images of a point source) to infer substructure effects and constrain dark matter mass models \citep{gilman_FRA, Cohen2024}.

The necessity for complex mass modelling has gained further support from recent studies with high-quality lensing data from instruments like ALMA \citep{2022MNRAS.516.1808P, stacey2024complex}, which recommend the inclusion of third and fourth-order multipoles in lens models. \citealt{arisM1} examined the stellar distribution in nearby massive elliptical galaxies, finding evidence for third and fourth-order multipoles. They also emphasized that in some cases, a first-order multipole (indicating lopsidedness in mass distribution) is crucial for accurate modeling, as seen in a recent detailed strong lens analysis survey \citep{barone2024lens}. Other studies have highlighted various forms of angular complexity, such as radial ellipticity variations and twists in the mass distribution\citep{nightingale2019decomposed, He2024}, and showed that lensing mass profiles cannot always be adequately described by simple power-law density profiles \citep{Etherington2023bulge}. Additionally, `External Shear' often used in lens modelling can account for some complexities that may be absent in the central lens mass model \citep{keeton97shear, Witt97shear, Cao2021, etherington2023strong}.

JWST \textit{NIRCam} imaging provides an unprecedented multi-wavelength view of strong lenses \citep[e.g.][]{rigby2023templates, bergamini2023glass, mercier2023cosmosring, vanDokkum2024cosmosring}. This capability is essential for testing lens models since the source morphology varies across wavelengths \citep[e.g.][]{arisBAR}, producing different lensed images, while the mass model and instrumental effects remain wavelength-independent. In this work, we perform independent lens model fits across multiple wavelengths for two strong lenses to assess: (i) whether angular mass complexity (in the form of multipoles) is consistently required for the lensing galaxies across wavelengths, and (ii) whether a dark matter substructure is consistently favored in each waveband. This multi-wavelength approach allows a detailed exploration of the degeneracy between these two model components.

The paper is structured as follows. In \S\ref{data} we describe the JWST data and the context of the systems studied. In \S\ref{method} we discuss our {\tt PyAutoLens} modelling pipeline and the details of our models. We also introduce the multipole perturbations and discuss how these and real subhalos are considered in the data. In \S\ref{Results} we show our model results for the multipole and subhalo fits, first to SPT2147, and then to SPT0418. In \S\ref{discussion} we discuss the results addressing the question of whether SPT2147 actually contains multipoles and/or a substructure. We summarise our conclusions in \S\ref{conclusions}. Throughout this work we assume the Planck 2015 cosmological parameters \citep{Planck2016} unless otherwise specified. Additionally we make use of both natural logarithm (denoted simply `$\log$') and log with base 10 (denoted specifically as `$\log_{10}$').

\section{The Data}\label{data}

We study two strong gravitational lenses: SPT2147-50 and SPT0418-47. We use JWST NIRCAM imaging from filters f277W, f356W and f444W (2.77, 3.56 and 4.44 microns respectively) for SPT2147 and f356W, f444W for SPT0418 - other (bluer) wavelength images exist, but lack a sufficient source signal-to-noise ratio (SNR) to produce informative models.  The lenses were imaged as part of the {\it TEMPLATES} survey (PI: Rigby, \citealt{rigby2023templates}). ALMA data is also available for these lenses and the source galaxy of SPT2147 has been studied in \citet{arisBAR}, with ALMA substructure analysis of both lenses in ALMA to be released in a subsequent work.

\begin{figure*}
    \centering
    \includegraphics[width=0.9\linewidth]{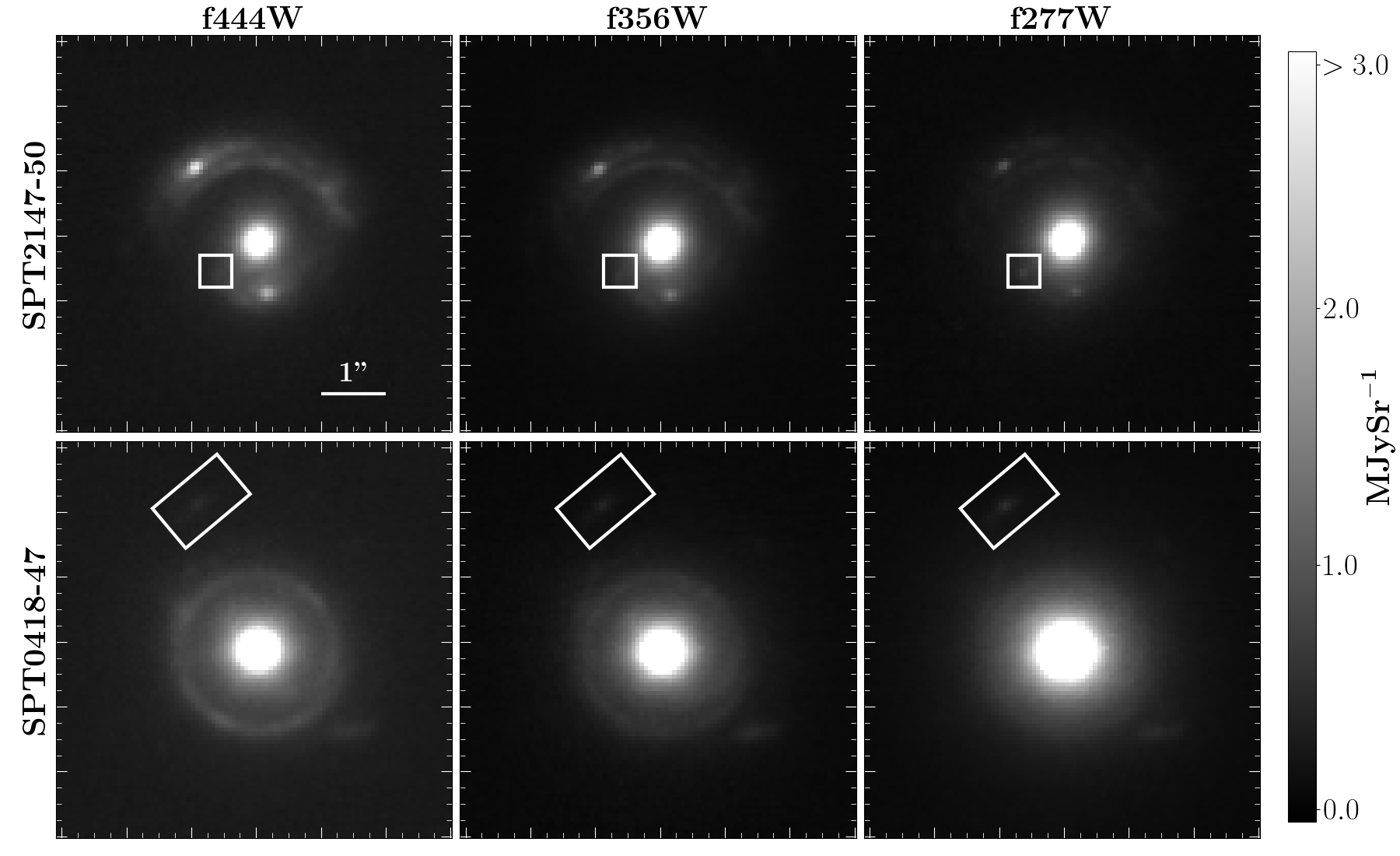}
    \caption{Data from the three reddest NIRCam filters (f277W, f356W, f444W) for SPT2147-50 (upper panels) and SPT0418-47 (lower panels). All panels are $6"$x $6"$ squares. The signal is saturated above 3.0 ${\rm MJy Sr}^{-1}$ in order to see the lensed source emission clearly. Marked with white boxes are line-of-sight galaxies in each lens (note the LOS object in SPT2147 is brightest in f277W).}
    \label{fig:data}
\end{figure*}

\begin{figure*}
    \centering
    \includegraphics[width=0.9\linewidth]{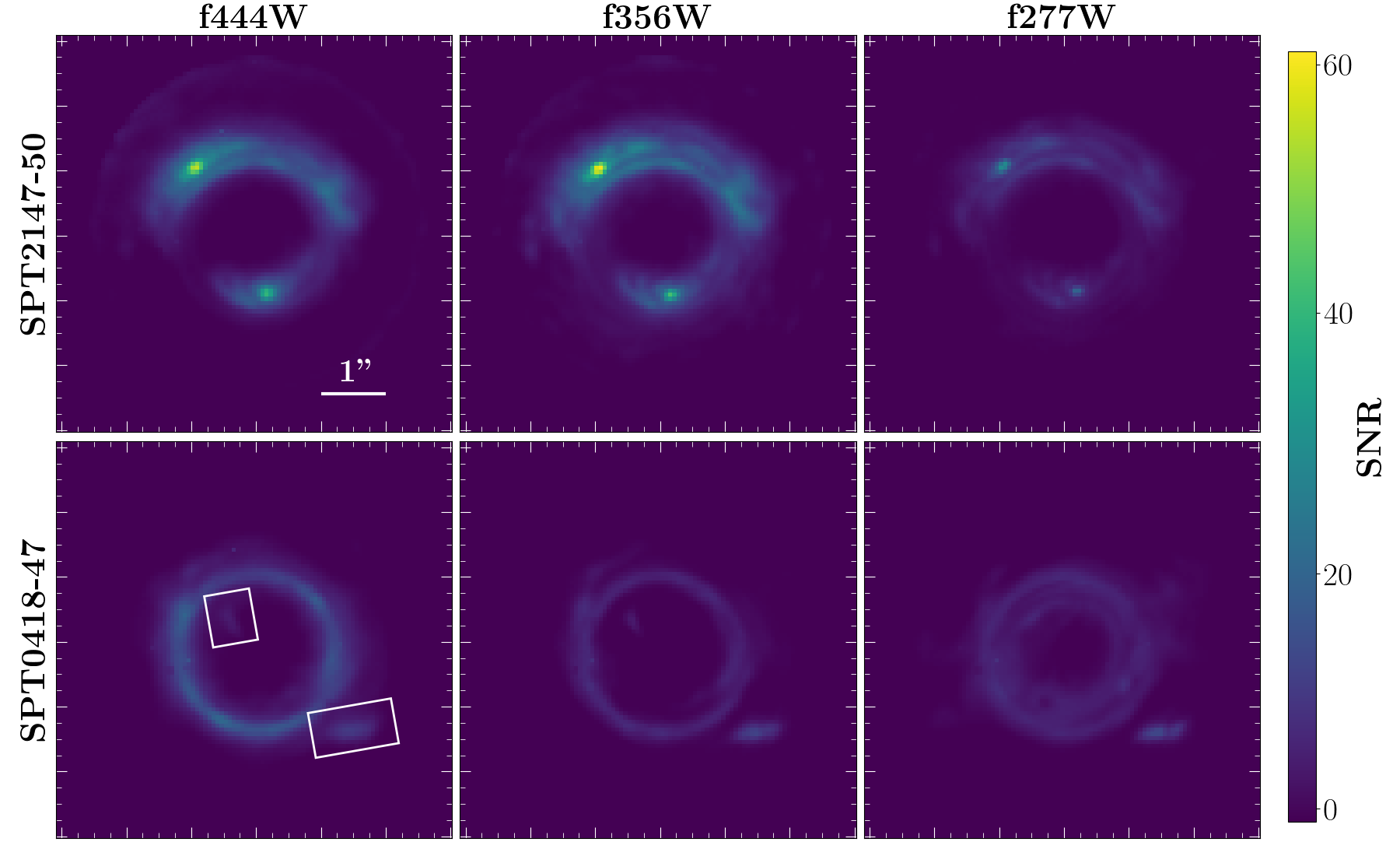}
    \caption{Signal-to-noise ratios (SNR) for SPT2147-50 (upper panels) and SPT0418-47 (lower panels), based on the model lensed source emission from our best-fit models. All panels are $6"$x $6"$ squares. All panels have color normalisation to the same level ([0,60]), but exact maximal and average values are shown in Table \ref{tab:SNR}. The white boxes in SPT0418 (f444W) mark the image and counter-image of the source companion galaxy as this lens has a multi-object source.}
    \label{fig:SNR}
\end{figure*}

JWST-NIRCam imaging is shown in Figure \ref{fig:data} for SPT2147 (upper) and SPT0418 (lower), where we have saturated all pixels above 3.0 ${\rm MJy Sr}^{-1}$ in order to see the lensed source galaxy light, as it is strongly outshone by the lensing galaxy. Figure \ref{fig:SNR} shows signal-to-noise ratio (SNR) maps for only the source emission, based on the model image created by our best-fit models (see Section \ref{Results}). We have normalised the colours to the same values across both lenses and all filters, to visualise the large difference between the maximum pixel SNR of the two lenses, with SPT2147 ranging from 2x to 5x higher SNR, depending on the filter. Specific values for maximal and average (over all pixels with ${\rm SNR} > 3$) SNRs are quoted in Table \ref{tab:SNR}.

\begin{table}
    \centering
    \begin{tabular}{Lcccccc}
         & \multicolumn{3}{c}{\textbf{SPT0418}} & \multicolumn{3}{c}{\textbf{SPT2147}} \\
         & \textbf{f444} & \textbf{f356} & \textbf{f277} & \textbf{f444} & \textbf{f356} & \textbf{f277} \\
         \hline
    \textbf{Max. Source SNR ${\rm pix}^{-1}$} & 20.90 & 12.57 & 14.68 & 53.80 & 59.59 & 28.79 \\
    \textbf{Avg. Source SNR ${\rm pix}^{-1}$} & 7.67 & 5.20 & 4.65 & 9.65 & 9.61 & 5.93 \\
    \textbf{RMS Noise ${\rm MJy Sr}^{-1}{\rm pix}^{-1}$}  & 1.606 & 1.654 & 1.455 & 1.599 & 0.930 & 1.259 \\
    \end{tabular}
    \caption{Signal information for each filter of our two strong lenses. The source parameters are based on the lens-light subtracted model images and therefore approximations based on model results. The avg. source SNR was calculated as the mean of all pixels with ${\rm SNR}>3$. The RMS noise was calculated over all pixels.}
    \label{tab:SNR}
\end{table}

\subsection{SPT 2147-50}\label{data2147}

This system is composed of a lensing galaxy at redshift $z=0.845$ and the background source galaxy at $z=3.76$ \citep{weib2013redshifts, reuter2020spt_redshift}. In conjunction with this work, this lens was also studied using ALMA data taken in band 8 by \citet{arisBAR}, where we discover that the background source galaxy is a spiral-barred galaxy. There is also a line-of-sight object southwest of centre, on the lower arc, marked with the white box in Fig.~\ref{fig:data}. This object is only really visible in f277W, and not visible at all in redder filters. Modelling the Spectral Energy Distribution (SED) of this object (only visible in the NIRCam bands) using the {\tt BAGPIPES}\footnote{\url{https://github.com/ACCarnall/bagpipes}} SED fitting code \citep{bagpipes1} gives a much lower redshift than our lens galaxy, in the range $z=0.04$ to $z=0.10$, and a best-fit mass of the order $\log_{10}({\rm M/M_{\odot}})\sim6$.

\subsection{SPT 0418-47}

This system is composed of a lensing galaxy at  redshift $z=0.263$ and a background source at redshift $z=4.22$ \citep{weib2013redshifts, reuter2020spt_redshift}. From our modelling and the literature, the source here is composed of two merging galaxies, with \citet{cathey2023_0418} reporting a projected separation of $\sim 4.42$ kpc and a \textit{stellar} mass ratio of 4:1. Interestingly, the source companion galaxy is not visible in HST or ALMA and so has only been discovered through JWST NIRCam, MIRI and NIRSpec data \citep{peng2023_0418, cathey2023_0418}. This companion galaxy is visible in the south-east of the image, outside the Einstein ring, with its counter-image visible north-west of the centre only once the lens light has been removed. Both are marked with the white box in the f444 panel of Figure \ref{fig:SNR}. There is also a nearby or line-of-sight galaxy, marked by the white box in Figure \ref{fig:data}, which we mask out of our analysis as it is too far from the main system to have any appreciable gravitational effect.

\section{Methods}\label{method}

To perform all the modelling described in this paper we use the open source lens modelling package {\tt PyAutoLens}\footnote{\url{https://github.com/Jammy2211/PyAutoLens}} \citep{pyautolens, Nightingale2015, Nightingale2018} and its associated parent packages, including {\tt PyAutoFit}\footnote{\url{https://github.com/rhayes777/PyAutoFit}}, an extension which provides a statistical fitting and analysis framework to our lens modelling \citep{pyautofit} and {\tt PyAutoGalaxy}\footnote{\url{https://github.com/Jammy2211/PyAutoGalaxy}}, a package which includes implementations of all galaxy mass and light profiles which may be needed for modelling \citep{pyautogalaxy}. 

The lens modelling procedure is similar to previous works using this software (e.g. \citealt{etherington2022automated}) and is as follows, with parameterisations as described in Table~\ref{tab:modelling}, and model choices detailed in the subsections below:
\begin{enumerate}
    \item \textbf{Initialise} - create a basic model for all light and lens mass using parametric profiles.
    \item \textbf{Source Pixelization} - reconstruct the source galaxy onto a pixel-based grid in order to capture higher levels of complexity.
    \item \textbf{Lens Refinement} - re-model the lens light based on this new pixelized source, and fit some more complex lens mass model.
    \item \textbf{Optional: Angular Mass complexity} - fit some perturbation to increase the angular mass complexity of the model (see Section \ref{multipole_method})
    \item \textbf{Subhalo Search} - introduce a subhalo to the lens mass model and investigate (see Section \ref{subhalo_method}).
\end{enumerate}

For all modelling stages, we use the non-linear sampler \texttt{nautilus}\footnote{\url{https://github.com/johannesulf/nautilus}} \citep{nautilus}, which uses Importance Nested Sampling to build a Bayesian Posterior estimate. We also remind the reader that `$\log$' refers to the natural log, and e.g. `$\log_{10}$' refers to the base-10 logarithm.

\subsection{Standard Mass and Light profiles}

\begin{table*}
    \centering
    \begin{tabular}{ACCCCCC}
        \textbf{Stage:} & \textbf{Initialise} & \textbf{Source Reconstruction} & \textbf{Lens Light Refinement} & \textbf{Lens Main Mass Refinement} & \textbf{Multipole perturbation} & \textbf{Subhalo Search} \\
        \hline
        \textbf{Source Light} & MGE: 1x30 & Voronoi mesh with Natural Neighbour interpolation & FIX model: linear solve only & FIX model: linear solve only & FIX model: linear solve only & FIX model: linear solve only  \\
        \hline
        \textbf{Lens Light} & MGE: 2x30 + 1x10 (+1x30 for SPT0418) & FIX model: linear solve only & NEW MGE: 2x30 + 1x10 (+1x30 for SPT0418) & FIX model: linear solve only & FIX model: linear solve only & FIX model: linear solve only \\
        \hline
        \textbf{Lens Mass} & Singular Isothermal Ellipsoid + External Shear & SIE + Shear (priors from previous) & FIX model: linear solve only & Elliptical Power-Law + New Shear & EPL+Shear (priors from previous) + Multipoles & EPL+Shear + Multipoles (priors from previous) + Spherical NFW \\
        \hline
    \end{tabular}
    \caption{Brief overview of the model parameterisations at the various stages of our fitting pipeline.}
    \label{tab:modelling}
\end{table*}

All light and mass profile quantities are computed with regards to their 'elliptical components' 
\begin{equation}
\begin{split}
& \epsilon_{1} =\frac{1-q}{1+q} \sin 2\phi, \\ 
& \epsilon_{2} =\frac{1-q}{1+q} \cos 2\phi,  
\label{eqn: ell_comps}
\end{split}
\end{equation}
where $q$ is the axis-ratio and $\phi$ the anti-clockwise position angle, defined as $\phi = 0$ when aligned to the positive x-axis. 

\subsubsection{Lens Mass: Elliptical Power-Law}

We model the lens mass through the elliptical power-law (EPL) density profile \citep{tessore2015EPL}, which is commonly used in strong lens modelling \citep[e.g.][]{nightingale2023scanning, vegetti2012detection}. This profile represents the total mass profile of the lens in the following form
\begin{equation}
\label{eqn:EPL}
\kappa (\xi) = \frac{(3 - \gamma^{\rm lens})}{1 + q^{\rm lens}} \bigg( \frac{\theta^{\rm lens}_{\rm E}}{\xi} \bigg)^{\gamma^{\rm lens} - 1} ,
\end{equation}
where the superscript `lens' implies the parameters relating to the primary lens mass (i.e. the `Macro-model', not any additional subhalo structures). Here, $q$ is the minor to major axis-ratio and $\xi$ is the elliptical coordinate defined as $\xi = \sqrt{{x}^2 + y^2/q^2}$. Additionally, $\gamma^{\rm lens}$ is the density slope, and $\theta^{\rm lens}_{\rm E}$ is the Einstein Radius (in arcseconds). $\gamma=2.0$ gives a Singular Isothermal Ellipsoid (SIE) which is used to initialise the lens mass model.

\subsubsection{External Shear}

An external shear is included in all stages of modelling. Some investigations note that there can be strong degeneracy of the shear with other model parameters (e.g. power-law ellipticity, \citealt{johnson2024foreground}) and there is evidence that the shear may account for missing internal mass complexity in the lens \citep{etherington2023strong}. 
We fit the shear as two elliptical components $(\gamma_{\rm 1}, \gamma_{\rm 2})$, and akin to the elliptical components of the main galaxy EPL, we relate the shear components to the magnitude, $\gamma_{\rm ext}$ and position angle $\phi_{\rm ext}$ (again, with the angle defined counter-clockwise from the positive x-axis):

\begin{equation}
    \begin{split}
    \label{eqn: shear}
    & \gamma^{\rm ext} = \sqrt{\gamma_{\rm 1}^{\rm ext^{2}}+\gamma_{\rm 2}^{\rm ext^{2}}}, \\
    & \tan{2\phi^{\rm ext}} = \frac{\gamma_{\rm 2}^{\rm ext}}{\gamma_{\rm 1}^{\rm ext}}.
    \end{split}
\end{equation}

\subsubsection{Multi-Gaussian Expansion}

To initialise the source galaxy light, and to fit the lens galaxy light, we apply a combined set of 2D elliptical Gaussian light profiles, known as a Multi-Gaussian Expansion \citep[][ hereafter MGE]{MGEs_cap}, used instead of the usual elliptical S\'ersic profile to attain increased model flexibility and accuracy, as described in \citet{He2024}. The intensity of an MGE set is given as
\begin{equation}
\label{eqn: MGE}
    I_{\rm set}(x, y) = \sum_{i}^{N} G_i(x, y).
\end{equation}
with $G_i$ the $i$-th Gaussian Profile,
\begin{equation}
\label{eqn: gaussian}
    G_i(x, y) = I_{i}\cdot{\rm exp}\left(-\frac{R_i^2(x, y)}{2\sigma_i^2}\right),
\end{equation}
containing $I_i$ as some intensity normalisation factor and $\sigma_i$ as the full-width at half-maximum of the gaussian profile. $R_i(x, y)$ is then the Gaussian elliptical radius, and given as
\begin{equation}
    \begin{split}
        & R_i(x, y) = \sqrt{x^{\prime2} + \left(\frac{y^\prime}{q_i}\right)^2} \\
        & x^{\prime} = \cos{\phi_i}\cdot\left(x - x^{\rm c}_i\right) + \sin{\phi_i}\cdot\left(y - y^{\rm c}_i\right) \\
        & y^{\prime} = \cos{\phi_i}\cdot\left(y - y^{\rm c}_i\right) -\sin{\phi_i}\cdot\left(x - x^{\rm c}_i\right),
    \end{split}
    \label{eq: ell_rad}
\end{equation}
where $q_i$ is the axis ratio, $\phi_i$ the position angle, and $\left(x^{\rm c}_i,\ y^{\rm c}_i\right)$ is the centre of the Gaussian. A "set" of Gaussians is used to refer to $n$ Gaussians which share the same axis ratio and position angle, but have sigma values increasing in fixed $\log_{10}$ intervals. We do not allow any individual Gaussian intensity to be negative, in order to avoid lens-source light degeneracy, solved using a modified\footnote{The fnnls code we are using is modified from \url{https://github.com/jvendrow/fnnls}.} fast non-negative least squares (fnnls) algorithm \citep{Bro1997MGEs}.

We combine 30 Gaussians for the source light initialisation. The number of gaussians used for the lens light varies between our two lenses; for SPT2147, we use 2x sets of 30 gaussians (akin to a bulge and disk component) plus 1x set of 10 gaussians to model central point-like emission. For SPT0418, the lens light is more blended with the lensed source emission and so we use 3x sets of 10 gaussians (akin to a thin disk, thick disk and bulge, or bulge-disk-envelope system), plus 1x set of 10 gaussians for central point-like emission. For both lenses, the centres of the sets of 30 gaussians is fixed to be the same (i.e. only one centre is fit for all), with the point-source set free to have a different centre, and the position angle/axis ratio of gaussians are the same throughout a set, but different between sets. The sigma values (controlling the `width' of the gaussian) are set between a lower limit of one-fifth the pixel scale (here $0.063"/5 = 0.0126"$) and an upper limit of 3 arcseconds for the main bulge/disk sets, and 0.126 arcseconds (2 times the pixel scale) for the point-like sets.

\subsubsection{Source Light: Voronoi Mesh Reconstruction}

The full formalism of the source reconstruction described and tested in detail with regards to substructure investigations on HST imaging of strong lenses in Appendix A of \citet{He2024} and in \citet{nightingale2023scanning} and is based on \citet{nightingale2015adaptive}. In general, for a given set of mass model parameters, once we subtract the lens light, the remaining lensed source emission is ray-traced back to the source plane and reconstructed onto an adaptive mesh "pixel grid". We use the Voronoi mesh grid, with a Natural Neighbour interpolation\footnote{More details about the natural neighbour interpolation technique can be found at \url{https://gwlucastrig.github.io/TinfourDocs/NaturalNeighborTinfourAlgorithm/index.html}.} \citep{Sibson1981} and our own adaptive regularisation to smooth the reconstruction based on the source luminosity at that point\footnote{In {\tt PyAutoLens} these schemes are the {\tt VoronoiNNBrightnessImage} pixelization and the {\tt AdaptiveBrightnessSplit} regularization.}. 

Equations \ref{eq:lensing_mge_goodness} describe the numerical implementation (without intermediate steps) of the "goodness of fit" function which optimises the balance between the MGE lens light and pixelised source light, allowing both to be fit simultaneously.

\begin{subequations}\label{eq:lensing_mge_goodness}
    \begin{align}
        G & = \chi^2 + G_{\rm S} + G_{\rm M} \label{equ: lensing_mge_goodness_l1} \\
          & = \frac{1}{2}\left(\sum_{i=1}^{N_{\rm I}}\left[\frac{\sum_{j=1}^{N_{\rm s}}f_{ij}S_{j} + \sum_{k=1}^{N_{\rm g}}I_kA_{ik} - d_i}{n_i}\right]^2 + G_{\rm L} + G_{\rm M}\right) \label{equ: lensing_mge_goodness_l2} \\
          & = \frac{1}{2}\left|\left|\left(
          \begin{array}{cc}
               Z & X \\
               \sqrt{r}B_S & B_{\rm M} 
          \end{array}
          \right)\cdot \left(
          \begin{array}{c}
               S  \\
               L
          \end{array}
          \right) - 
          \left(\begin{array}{c}
               Y  \\
               0 
          \end{array}\right)
          \right|\right|^2, \label{equ: lensing_mge_goodness_l3}
    \end{align}
\end{subequations}

Within equation \ref{equ: lensing_mge_goodness_l1}, $G_S$ and $G_M$ are regularisation terms for the pixelised source and the MGE light respectively. Equation \ref{equ: lensing_mge_goodness_l2} expands the chi-squared term in relation to the image-plane to source-plane mapping, where $S_j$ is the j-th source-plane pixel flux, $I_k$ is the k-th image-plane pixel flux, $f_{ij}$ is the mapping matrix element quantifying the contribution of source pixel $j$ to the flux of image pixel $k$ (which includes psf and image processing effects). In addition, the term includes $A_{ik}$, which is the exponential from the Gaussian profile of equation \ref{eqn: gaussian} $G_k$, for $(x_i, y_i)$, as well as the data and noise $d_i$ and $n_i$ at the i-th pixel.
All of this collapses down into equation \ref{equ: lensing_mge_goodness_l3}, where $Z_{ij} = f_{ij}/n_i$, $X_{ij}=A_{ij}/n_i$, $r$ is the regularisation strength, $B_S$ and $B_M$ are `square root' matrices of the source and MGE regularisations respectively, $S$ and $L$ are the source-plane and image-plane fluxes, and $Y_i$ is the division $d_i/n_i$.

\subsection{Angular Mass Complexity: Multipoles}\label{multipole_method}

To introduce additional complexity to the `macro-model', we fit an extension to the EPL profile where we include internal multipoles, building on the formalism of \citet{chu2013multipole}. An individual multipole perturbation adds to the EPL to give a functional form of the convergence, in polar coordinates, of: 
\begin{equation}
    \kappa(r, \phi) = \frac{1}{2} \left(\frac{\theta_{\rm E}^{\rm lens}}{r}\right)^{\gamma^{\rm lens} - 1}\cdot k_m \, \cos(m(\phi - \phi_m)) \, ,
\end{equation}
where $r$ is the radial position in arcseconds, $\phi$ is the angle of interest for the calculation (in degrees), and $m$ is the defined multipole order. 
When modelling, the Einstein Radius, $\theta^{\rm lens}_{\rm E}$, slope, $\gamma^{\rm lens}$ and $(y,x)$ centre of the multipole are linked to the values fit to the main EPL, and in addition, we fit for the multipole elliptcial components $(\epsilon_{\rm 1}^{\rm mp},\ \epsilon_{\rm 2}^{\rm mp})$ (meaning we have the six EPL parameters plus an additional two per multipole). The multipole elliptical components relate to the multipole strength, $k_m$ and multipole position angle $\phi_m$ (defined counter-clockwise from the positive $x$-axis) in a similar way to the EPL elliptical components, through 
\begin{equation}
\begin{split}
    \label{eqn: ell_comps_multipole}
    & \phi_m = \frac{1}{m}\arctan{\frac{\epsilon_{\rm 2}^{\rm mp}}{\epsilon_{\rm 1}^{\rm mp}}}, \\
     & k_m = \sqrt{{\epsilon_{\rm 1}^{\rm mp}}^2 + {\epsilon_{\rm 2}^{\rm mp}}^2}.
    \end{split}
\end{equation}

The deflection angles can also be given in polar co-ordinates (where $\alpha$ is the power-law slope) as: 
\begin{equation}
\begin{split}
\label{eqn: deflection_multipole}
    & a_r(r, \phi) = - \frac{3 - \alpha}{m^2 - (3 - \alpha)} r_E^{(\alpha - 1)} r^{(2 - \alpha)} k_m \cos\left[m(\phi - \phi_m)\right], \\
    & a_{\phi}(r, \phi) = \frac{m^2}{m^2 - (3 - \alpha)} r_E^{(\alpha - 1)} r^{(2 - \alpha)} k_m \sin\left[m (\phi - \phi_m)\right]. 
\end{split}
\end{equation}

Multipole perturbations cause the mass distribution to lose angular symmetry, but maintain rotational symmetry at different orders, represented by $m$ (i.e. $m=1$ has one line of symmetry, $m=2$ as two lines of symmetry, etc) and so in theory, any number of multipole perturbations could be added to the mass model, however, we restrict ourselves to three orders, $m=1$, $m=3$ and $m=4$. For this work, we test both individual perturbations as well as combinations, in which case the perturbations sum together, giving 

\begin{equation}
    \label{eqn: multipole}
    \kappa(r, \phi) = \frac{1}{2} \left(\frac{\theta_{\rm E}^{\rm lens}}{r}\right)^{\gamma^{\rm lens} - 1}\cdot \sum_{m} \bigg( k_m \, \cos(m(\phi - \phi_m))\bigg) \, .
\end{equation}

Some physical motivations and examples of the effects of multipole perturbations to lensing systems are shown in Appendix \ref{app:mulitpoles}.

\subsection{Subhalo Modelling and Bayesian Evidence}\label{subhalo_method}

We use a Spherical Navarro-Frenk-White (NFW) profile \citep{NFW} to model DM-only subhalo additions, and only model subhalos assuming the same redshift as the primary lensing galaxy. The density of an NFW with normalisation $\rho_{\rm s}$ and scale radius $r_{\rm s}$ is given as 
\begin{equation}
    \rho_{\rm NFW} = \frac{\rho_{\rm s}}{(r/r_{\rm s}) (1 + r/r_{\rm s})^2}.
    \label{eqn: NFW}
\end{equation}

We fit only for the $(x,y)$ centre of the subhalo and the $M_{200}$ mass (the enclosed mass at $r_{200}$, which is the radius at which the average enclosed density is 200 times $\rho_{\rm crit}$ for the Universe) and relate $M_{200}$ to the scale radius through the mean mass-concentration relation given by \citet{ludlow2016mcr}.

Lens models including a Dark Substructure produce complex multi-modal parameter spaces which are challenging for non-linear tools such as Nautilus to sample robustly. We therefore split the image into a 5x5 grid (each 'pixel' being 1 square arcsecond) and model a subhalo where the positional priors are within that grid cell only \citep[see e.g.][]{nightingale2023scanning}. Following this, we can both visualize the effect of adding a subhalo at different locations, as well as use the result from the grid cell with the highest Bayesian Evidence increase to initialise a new fit for a subhalo where the model has the freedom to place the subhalo anywhere, but the positional priors are gaussians centred on the best-fit position of the highest-evidence grid search.

\subsubsection{Model Discriminators: Bayesian Evidence}\label{evidence_method}

We use the natural ${\rm log}$ Bayesian Evidence - \lev - as the discriminating factor between models. A simplified way to understand the Bayesian Evidence is to consider it to be the Likelihood of the Model being the truth, but marginalised over the parameters that are being fit, such that say a small Prior which contains many bad solutions will have a lower evidence than a larger Prior that only contains good solutions. The equation
\begin{equation}
    \mathlarger{\mathlarger{\mathlarger{\varepsilon}}} = p(\mathbf{X}|\alpha) =  \int p(\mathbf{X}|\theta)p(\theta |\alpha) d\theta
    \label{eqn: Evidence}
\end{equation}
formalizes this description, where $\mathbf{X}$ is some observed data, $\theta$ is a data parameter, $\alpha$ is a prior hyperparameter, such that the evidence is $p(\mathbf{X}|\alpha)$ (probability of data given prior parameters), which is then the integral over all data parameters of the Likelihood ($p(\mathbf{X}|\theta)$, probability of data given a parameter set) multiplied by the Prior ($p(\theta |\alpha)$, probability of a certain data parameter set given the prior hyperparameter sets).

To claim a subhalo candidate against a standard EPL mass model, we would normally set an a-priori requirement of \dlev $ > 10$ for the subhalo addition, and we set the same standard for the inclusion of multipoles in a model. For reference, \dlev $= 10$ corresponds to roughly a 5-sigma result\footnote{A useful online source for Bayes Factors in context can be found here: \url{https://ned.ipac.caltech.edu/level5/Sept13/Trotta/Trotta4.html}}, or a frequentist p-value of order $10^{-7}$. Similarly, \dlev $ = 5$ corresponds to roughly 3-sigma or a p-value of $\sim 0.03$. The use of \dlev can also be considered as the logarithm of the `Bayes Factor' or `Odds', since \dlev $ = {\rm log}(\mathlarger{\mathlarger{\mathlarger{\varepsilon}}}_1) - {\rm log}(\mathlarger{\mathlarger{\mathlarger{\varepsilon}}}_2) = {\rm log}(\mathlarger{\mathlarger{\mathlarger{\varepsilon}}}_1/\mathlarger{\mathlarger{\mathlarger{\varepsilon}}}_2)$.

Our specific implementation of evidence thresholds in this work is to hold the requirement of \dlev $> 10$ when we add multipoles to the EPL+Shear model, and if these multipole-inclusive models pass the threshold then we continue to add a subhalo to the model. If the multipole evidence is not satisfied, we would only apply the subhalo to the base model\footnote{We add subhalos to some models in SPT0418 despite them not passing the set evidence threshold only to attain a more thorough comparison to SPT2147.}. 

When using only single-filter imaging, if there are many different combinations of multipoles and multipoles+subhalos that create an evidence change of \dlev $> 10$ to the base model, it would be difficult to decide between taking just the highest evidence model, or requiring that a subhalo has a further evidence to the multipole (due to unknown degeneracies between multipoles and subhalos). Since we have multi-filter fitting available, we are able to apply the `highest evidence model' approach and cross-reference by applying a \dlev $= 5$ (3$\sigma$) confidence region below this model - comparing results which fit into the confidence region across all filters. If more than one model was within this region, we choose the highest evidence model from across all filters as the primary candidate (i.e. the highest evidence model from f444W, since this has the highest source SNR).

\section{Results}\label{Results}

\subsection{SPT2147 Results}\label{results_2147}

The SED of the line-of-sight object to the southwest of centre on the lower arc was modelled with {\tt BAGPIPES}, as mentioned in Section \ref{data2147}, to obtain $0.04 \lesssim z \lesssim 0.10$, and a best-fit mass of the order $\log_{10}({\rm M/M_{\odot}})\sim6$. We initially try to include this object in our lens modelling, however the einstein radius converges to zero in all fits, meaning the object has no appreciable lensing effect in the system. We therefore remove the object from our subsequent fits by artificially increasing the noise level over this area, so that the light at this point provides no contribution to our likelihood functions, due to the now very low signal-to-noise.

\subsubsection{Base model substructure results}\label{2147base}

\begin{figure*}
    \centering
    \includegraphics[width=0.99\linewidth]{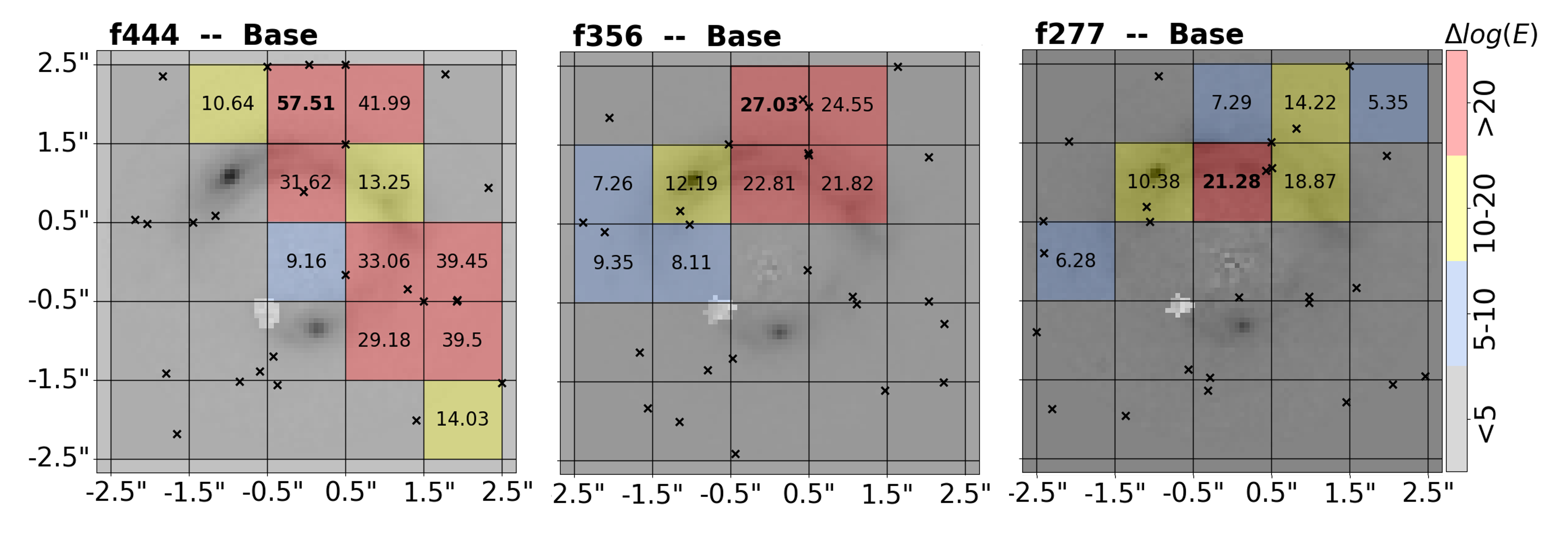}
    \caption{Results from the grid-based preliminary subhalo searches in SPT2147 for the base EPL+Shear model. The values and colours correspond to the Log-Evidence change - \dlev - when a subhalo is modelled in a certain grid cell. Crosses mark the best fit position of the subhalo within each grid cell. The highest evidence grid-cells are marked in bold. \dlev $ = 10$ is considered a roughly 5$\sigma$ detection.}
    \label{fig:2147grid_base}
\end{figure*}

We show in Figure \ref{fig:2147grid_base} our results from the subhalo grid-search analysis (see Section \ref{subhalo_method}), in each filter, using the EPL + shear as the base macromodel. This `Base' model shows multiple high-evidence (\dlev $> 10$) grid squares repeated in similar locations in each filter. The subhalo's location with the highest evidence is north of the extended arc, with maximal evidence increases of $\sim 57, 27, 21$ in the grid-search for f444, f356 and f277 (note that these values adjust to $60.35, 42.13, 20.54$ for the final DM subhalo fit after the grid search, where the priors on its centre are not confined to a 1x1 arcsecond square).

The f444 filter also gives a secondary set of high-evidence grid-cells, east of the lensing galaxy, which are not seen in the bluer filters. The reason these right-hand cells do not show high evidences in the f356 and f277 filters may be related to the much lower source SNR in f277, and potentially the source light morphology in f356. We have seen from simulations that genuine subhalo signals can also produce residual signal in incorrect locations \citep[][section B2]{nightingale2023scanning}, so this could just be a residual effect of a genuine subhalo located towards the top of the image. There may also be a case where the substructure signals in f444 are not due to a genuine substructure, but rather a product of missing complexity from the EPL+Shear in that area, where the model uses the inclusion of substructure to compensate for this - the combination of any slight model differences with source morphology changes between filters could then also explain why these are not seen in f356/f277.

\subsubsection{Results of Multipole Perturbations}\label{2147mp}

\begin{figure}
    \centering
    \includegraphics[width=0.99\linewidth]{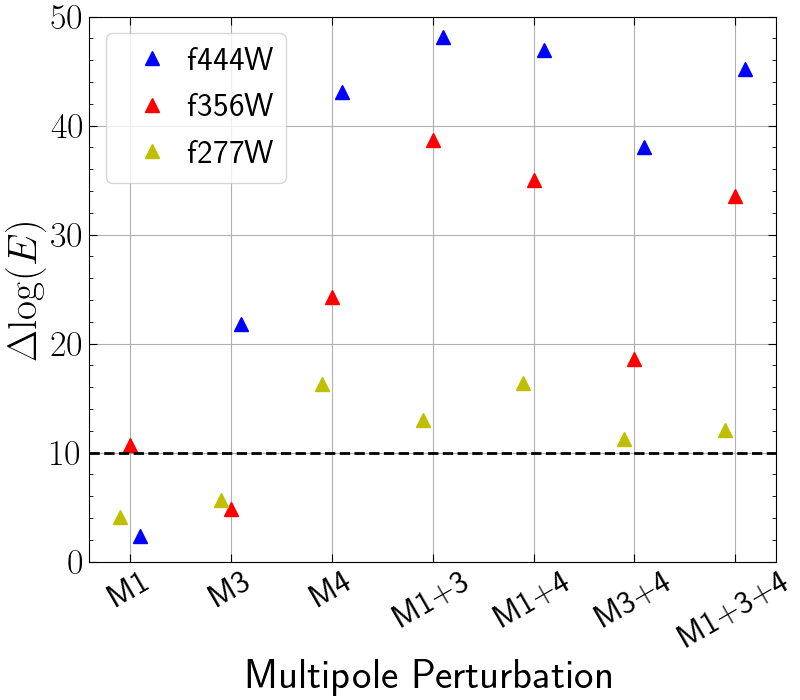}
    \caption{\dlev (Log-Evidence changes) with the addition of multipole perturbations to the base model of SPT2147. The dashed line marks \dlev $=10$, which is our approximate 5$\sigma$ significance threshold.}
    \label{fig:2147mp_evi}
\end{figure}

\begin{figure*}
    \centering
    \includegraphics[width=0.99\linewidth]{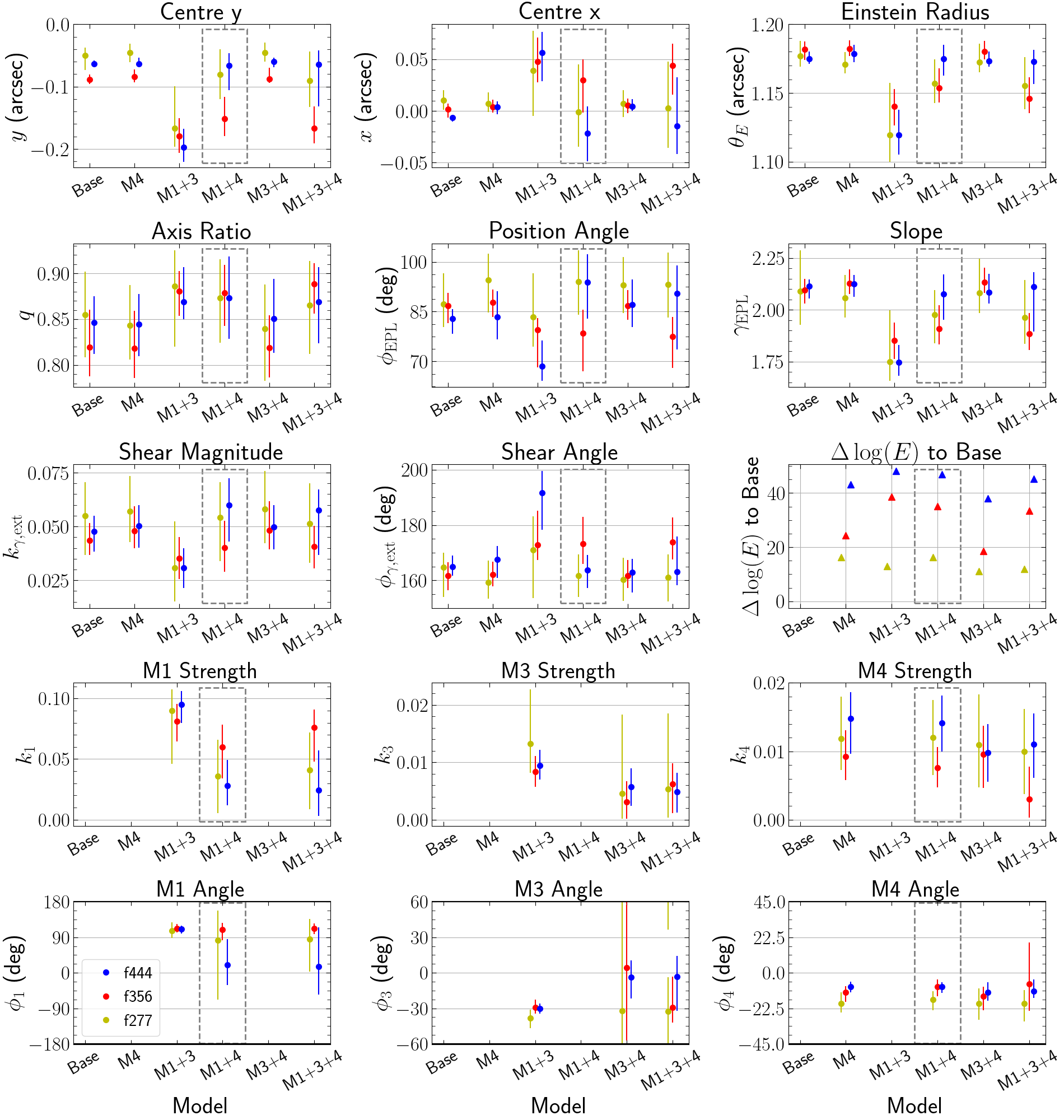}
    \caption{Reported parameters and associated 3$\sigma$ errors for the highest evidence models without a subhalo in SPT2147, and the Base model. Multipole angles have a symmetry through $360/n$ - i.e. a 90 degree error on the M4 multipole means it is unconstrained as this multipole has rotational symmetry through 90 degrees, and so the errors are wrapped around $\pm (360/n)/2$. Evidence changes shown are in comparison to the Base EPL+Shear model. We highlight with the gray box our preferred final model of EPL+Shear+M1+M4.}
    \label{fig:2147mp_params}
\end{figure*}

Figure \ref{fig:2147mp_evi} shows the log-evidence increases - \dlev - when we include various multipole perturbations to our EPL+Shear model, instead of the subhalo additions from the previous section. We see that all multipole inclusions are preferred (give a positive evidence change), and all apart from M1-only and M3-only give \dlev $> 10$ in all three filters. From here forwards therefore, we discount the M1-only and M3-only fits unless otherwise stated, similarly discounting the f277 fits due to the low SNR (and therefore low sensitivity to model changes).

From Figure \ref{fig:2147mp_evi}, we also see that the M4 (in f356) and M3+4 (in f444 and f356) fits, show a lower evidence than the other multipole additions, which is a potential indication that, whilst an M1 multipole addition on its own cannot produce much difference to the base model, it may be required in order for the higher-order M3 and M4 perturbations to become fully effective - a possible sign again of multipoles having the potential to compensate for missing complexity which they are not primarily intended to capture. A similar effect is shown in \citet{arisM1} where M1 addition changes the fit values of M3 and M4 components to galaxy light distributions. 

Figure \ref{fig:2147mp_params} shows the recovered parameters from the high-evidence multipole models to SPT2147, as well as the base EPL+Shear model. For each model, the parameters are consistent within the errors across all three filters, which shows that the fits are converging successfully since, again, the three filters offer independent source light structures and PSF instances but with the same mass distribution, and thus agreement between the three filters within a model provides an important sanity check and confidence boost in the recovered parameters. Between different models however, we see one model notably deviate from the base model (and indeed most other model) parameters - the M1+M3 model, showing an offset centre, smaller Einstein Radius, smaller Power-Law Slope, smaller Shear Magnitude and higher M1 magnitude. Some of these M1+M3 parameters are consistent within errors to at least one filter of another model but the best-fit values are clearly offset, which is important as this model actually has the highest evidence of all the models in f444 (\dlev $=48.16$) and f356 (\dlev $=38.69$). In terms of result trends between multipoles, we do also see a relation between the M3 and M4 multipoles in this lens where the best-fit M3 magnitude slightly decreases and its angle becomes much less constrained once we add the M4 multipole (in M3+M4 and M1+M3+M4), and similarly for the M4 magnitude when the M3 is added. The M1 magnitude also shows decreases when the M4 is added compared to the M3.

\subsubsection{Inclusion of both Multipoles and Substructure}\label{2147combi}

\begin{figure*}
    \centering
    \includegraphics[width=0.81\linewidth]{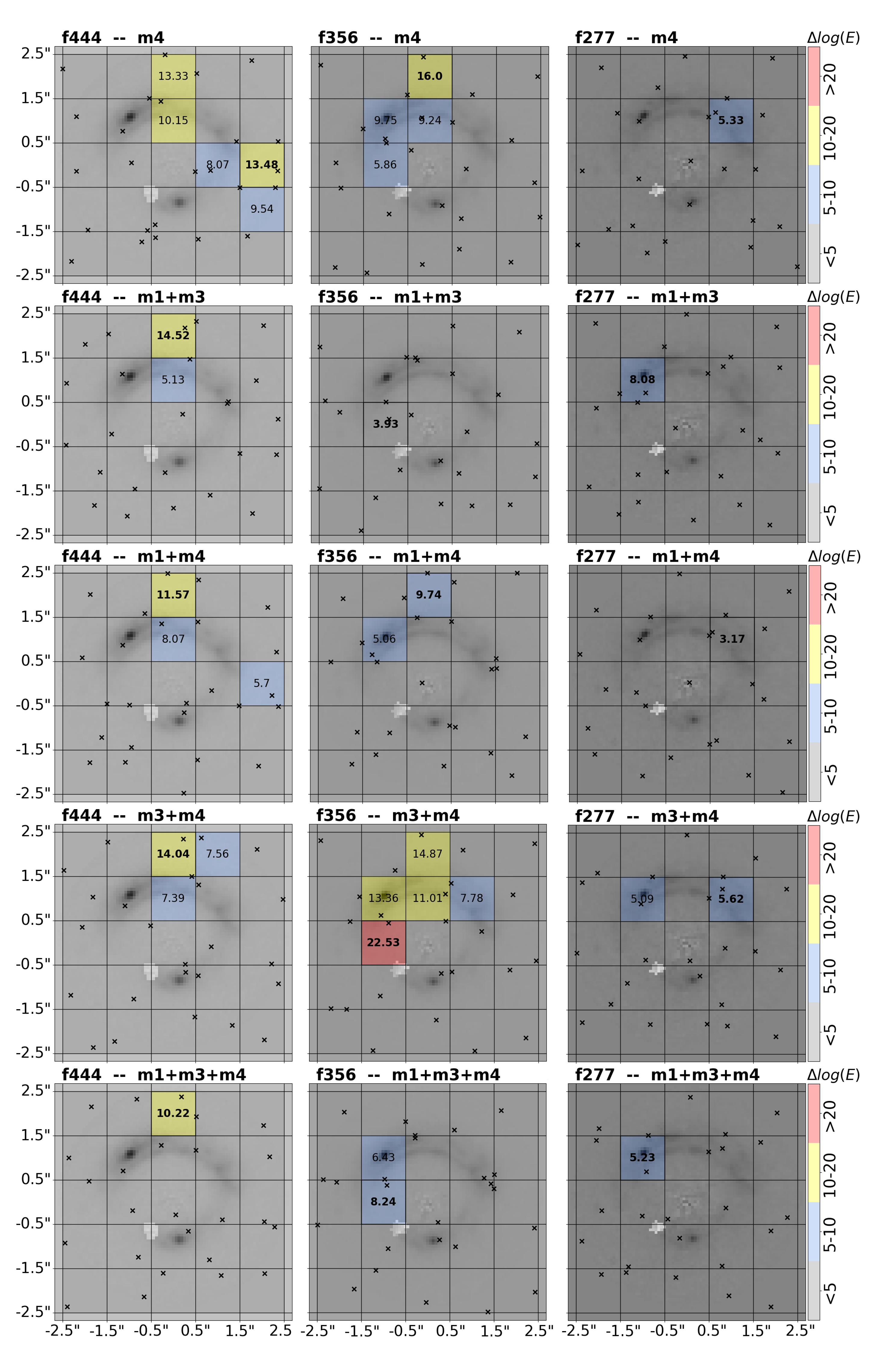}
    \caption{Results from the grid-based preliminary subhalo searches in SPT2147 for the EPL+Shear+Multipole models. The values and colours correspond to the Log-Evidence change - \dlev - when a subhalo is modelled in a certain grid cell. Crosses mark the best fit position of the subhalo within each grid cell. The highest evidence grid-cells are marked in bold. \dlev $= 10$ is considered a roughly 5$\sigma$ detection.}
    \label{fig:2147grid_mp}
\end{figure*}

\begin{figure*}
    \centering
    \includegraphics[width=0.99\linewidth]{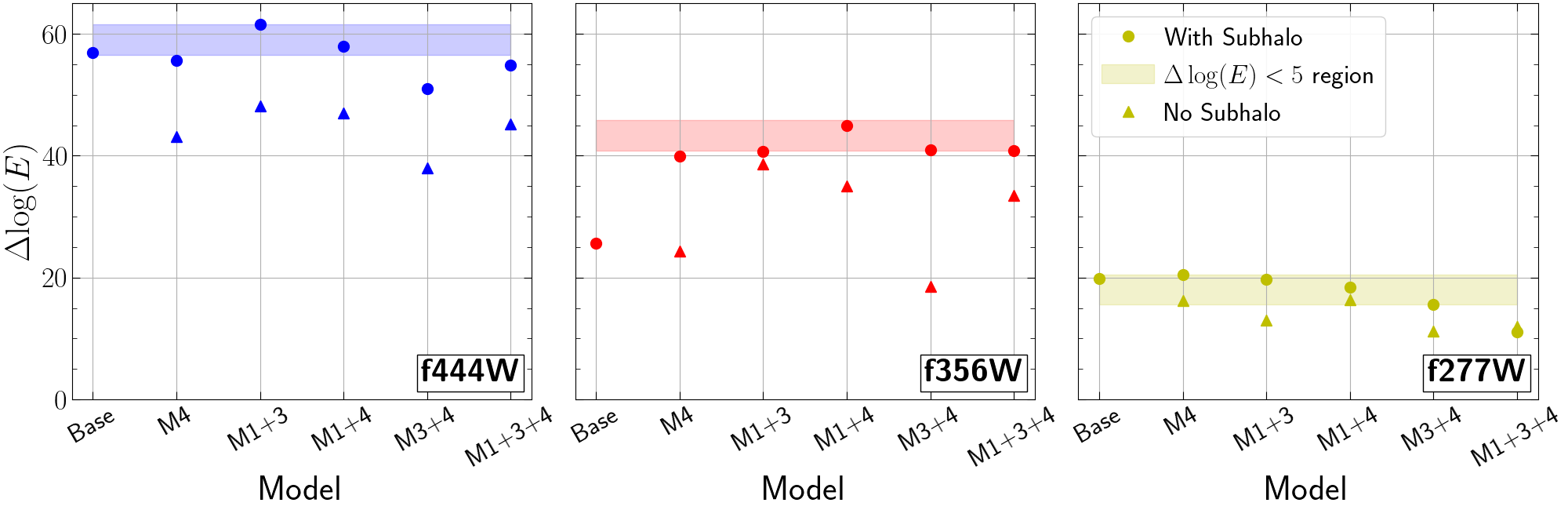}
    \caption{\dlev (Log-Evidence changes) for the addition of a multipole (triangles, for reference), and then subhalos (circles) to highest evidence models of SPT2147. All evidences are compared to the Base no-subhalo model. The shaded regions correspond to the (roughly 3$\sigma$) \dlev $<5$ confidence region below the highest evidence models in each filter.}
    \label{fig:2147evi_sub}
\end{figure*}

Evidence increases from the addition of multipoles to the EPL+Shear model, and the addition of substructures to this model, are very similar in magnitude, which again may be a possible indication that there is degeneracy between multipoles and substructure in terms of each compensating for the absence of the other, adding to the results of \citet{oriordan2023multipoles}. Running the subhalo searches on these models which have multipole perturbations allows us to see that both can actually co-exist. Figure \ref{fig:2147grid_mp} shows the subhalo grid search results for the multipole models, with every multipole model shown having a grid cell with \dlev $> 10$ in f444. No models show a cell with \dlev $ > 10$ in f277, but we do not draw strong conclusions from this filter due to the low source signal (and therefore low constraining power). The M3+M4 model has a high-evidence grid cell (\dlev $ > 20$) in f356, potentially accounting for the slight dip seen in the evidence for the multipole perturbation in this model. Importantly, most models in f444 and f356 (except M1+M3-f356 and M1+M3+M4-f356) still favour a subhalo in the same or similar position to the subhalo suggested from the base model search, showing that even with the addition of multipole perturbations, a substructure still obtains higher Bayesian Evidences for the model-fit.

Figure \ref{fig:2147evi_sub} shows the finalised multipole evidences and subhalo evidences for the base model and multipole models. We see two results here: firstly that in f444 the multipole perturbations alone cannot reach the same evidence increase that just adding a subhalo to the base model can, however in f356, the multipole perturbations can reach and surpass this level; second, we see that many models lie within a reasonable evidence range of one another, showing the difficulty we will have to choose the 'best' models confidently without informative physical priors and context. Following the method set out in Section \ref{evidence_method} however, we utilise the multi-wavelength fitting advantages of these data and plot the confidence region of \dlev $< 5$ below the highest overall evidence models in each filter as the shaded regions in Figure \ref{fig:2147evi_sub}. Again, neglecting the f277 filter due to its low constraining power, we see that the M1+M3+subhalo and M1+M4+subhalo models are the only models which lie within this region in both f444 and f356. Since M1+M3+subhalo in f356 lies both at the edge of the region and only \dlev $\sim 2$ above the M1+M3 model, we take M1+M4+subhalo as the preferred model. Additionally, the M4 multipole perturbation is more easily related to physical structure than the M3 multipole, providing another reason to take this as the preferred model. 

\begin{figure*}
    \centering
    \includegraphics[width=0.9\linewidth]{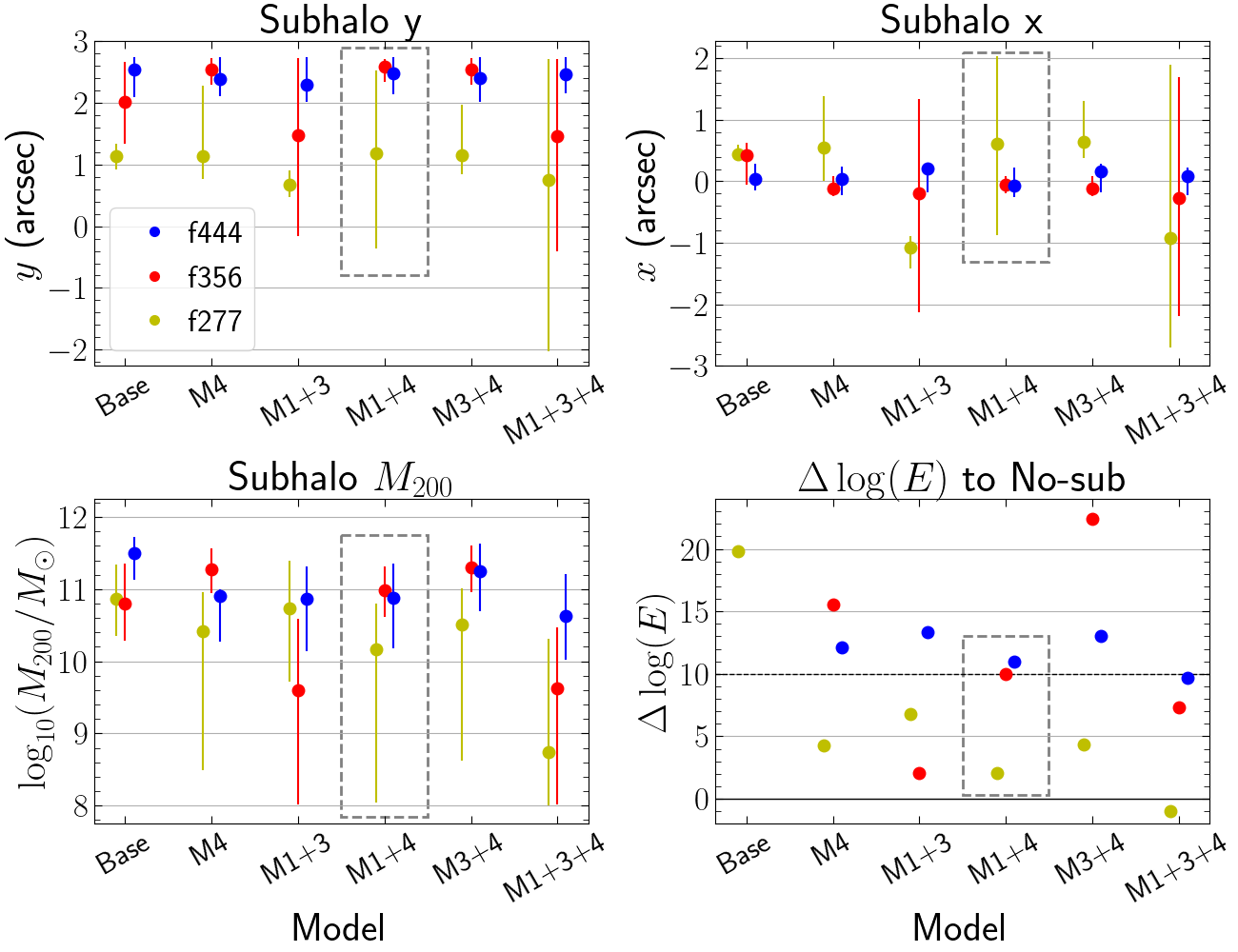}
    \caption{Reported subhalo parameters for the best-fit substructure to the Base and high-evidence multipole models. Our preferred model, EPL+Shear+M1+M4 is highlighted with the gray box. Log-evidence changes - \dlev - compared to the no-subhalo model of that type are also shown, but the Base f356 and f444 evidences are above the plot limit.}
    \label{fig:2147subhalo_params}
\end{figure*}

\bgroup
\def\arraystretch{1.5}
\begin{table}
    \centering
    \begin{tabular}{cccc}
         & \textbf{f444} & \textbf{f356} & \textbf{f277} \\
         \hline
    \dlev to Base & 57.96 & 45.00 & 18.46 \\
    \dlev to M1+4 (no-sub)  & 10.97 & 9.96 & 2.07 \\
    Position $y$ (arcsec) & $2.47^{+0.27}_{-0.33}$ & $2.59^{+0.13}_{-0.25}$ & $1.19^{+1.34}_{-1.55}$ \\
    Position $x$ (arcsec) & $-0.07^{+0.29}_{-0.19}$ & $-0.06^{+0.15}_{-0.13}$ & $0.61^{+1.43}_{-1.48}$ \\
    ${\rm log}_{10}(M_{200}/M_{\odot})$ & $10.87^{+0.53}_{-0.71}$ & $10.99^{+0.33}_{-0.37}$ & $10.17^{+0.64}_{-2.12}$ \\
    \end{tabular}
    \caption{\dlev (log-evidence changes) and subhalo parameters for our prime model of EPL+Shear+M1+M4+subhalo. Errors quoted are 3$\sigma$. Evidences are quoted to 'Base' - the EPL+Shear model, and to 'M1+M4 (no-sub)' - the EPL+Shear+M1+M4 model before adding the subhalo. All values are consistent across three filters.}
    \label{tab:2147_M14S}
\end{table}
\egroup

Table \ref{tab:2147_M14S} shows the evidences, positions and masses for the recovered substructure in our favoured model, and Figure \ref{fig:2147subhalo_params} shows the subhalo position and mass for all filters and models. Within individual models, all masses agree between filters, but there are some disagreements between f444 and f277 on position in the Base, M1+3 and M3+4 models. Between models, on the other hand, the f444 and f356 subhalo positions all agree, with some disagreements on the subhalo mass, which is to be expected from models which have different mass distributions in the macromodel. We will discuss in Section \ref{disc:pos and mass} the potential interpretations of the model-internal disagreements. 

The macromodel parameters are shown in Appendix \ref{app:additional_params}, Figure \ref{fig:2147_sh_params}, as well as the changes in those parameters from the no-subhalo to the with-subhalo models in Figure \ref{fig:2147_sh_changes}. Generally, most changes are consistent with 0, although for best-fit values we see a general drop in M4 strength, and that the addition of a subhalo tends to bring the power-law slope closer to Isothermal ($\gamma = 2$). The most notable changes upon addition of a subhalo are to the Einstein radius, with the base, M4 and M3+M4 models showing a definitve drop in Einstein radius in f356 and f444. For our preferred M1+M4 model however, the addition of the subhalo leads to no notable change in Einstein radius - there is a potential increase in EPL slope, decrease in shear magnitude and decrease in M4 strength, but all with errors making these changes consistent with 0.

\subsubsection{Model Residuals}\label{2147_residuals}

\begin{figure*}
    \centering
    \includegraphics[width=0.99\linewidth]{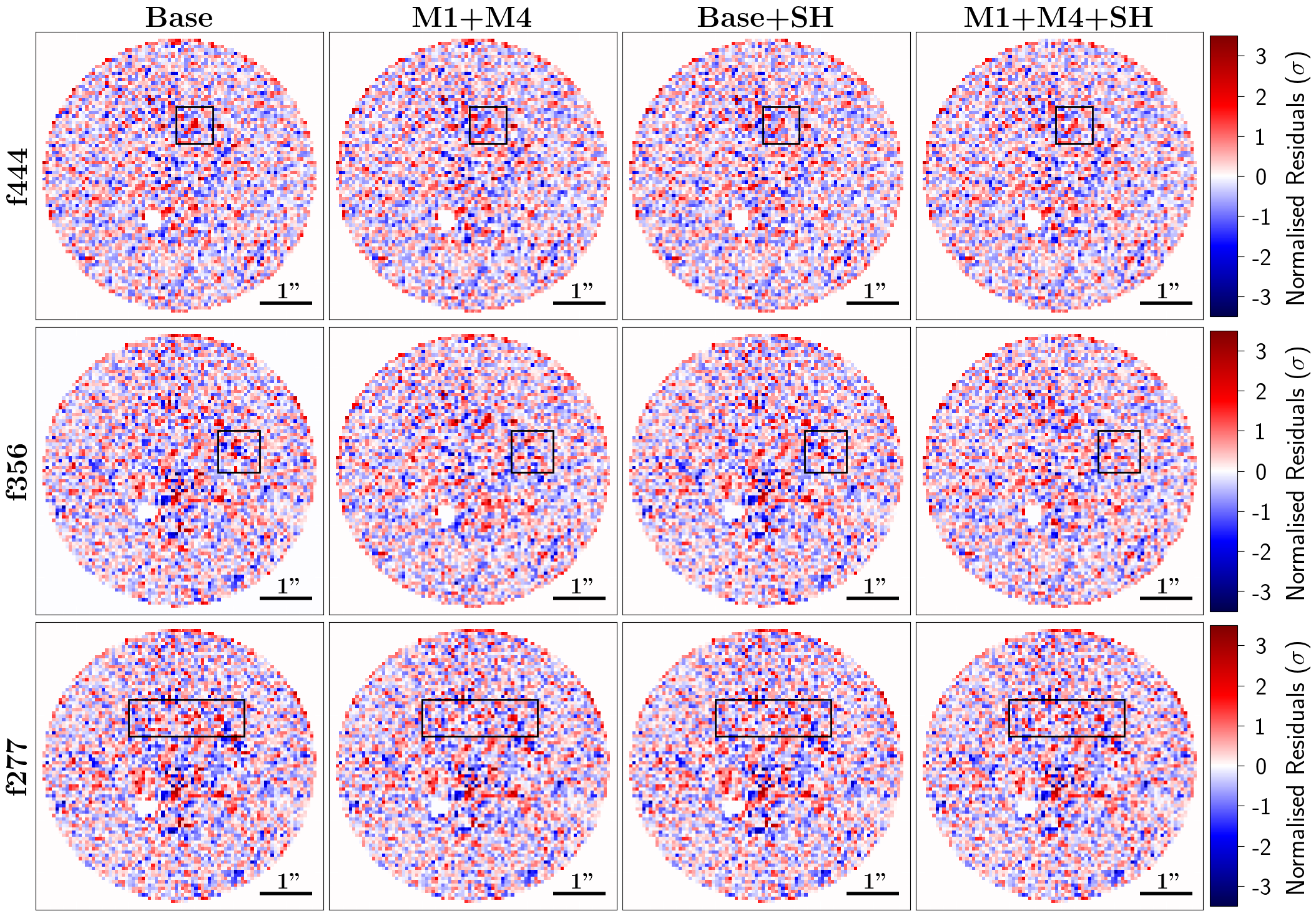}
    \caption{ Normalized residuals (residuals divided by the noise in each pixel) from model fits to SPT2147 of the `Base' (EPL+Shear) model and the model including $m=1$ and $m=4$ multipoles, and their corresponding fits with substructure additions (`SH'). The black boxes highlight areas where there are notable and correlated residual changes within a filter, all corresponding to the upper arc of the lensed source.}
    \label{fig:2147_residuals}
\end{figure*}

The normalized residuals (i.e. the model residuals divided by the noise, from here forward just `residuals') from the fits to SPT2147 of the Base, Base + Subhalo, Base + M1 + M4, and Base + M1 + M4 + subhalo models are shown in Figure \ref{fig:2147_residuals}. There are areas of the upper arc of the lensed source (marked with back boxes) which see noticeable changes in their residuals from the Base model to the model with multipoles and to the model with substructure, however, the change is much less obvious going from these models to the model with multipoles {\it and} substructure, apart from in the f356 filter, which is a good indicator that the increase in bayesian evidences we see is due to improvements in the modelling of the lensed source and not the lens itself or the noise.

\subsection{SPT0418 Results}\label{results_0418}

\subsubsection{Base model substructure results}\label{0418base}

\begin{figure*}
    \centering
    \includegraphics[width=0.9\linewidth]{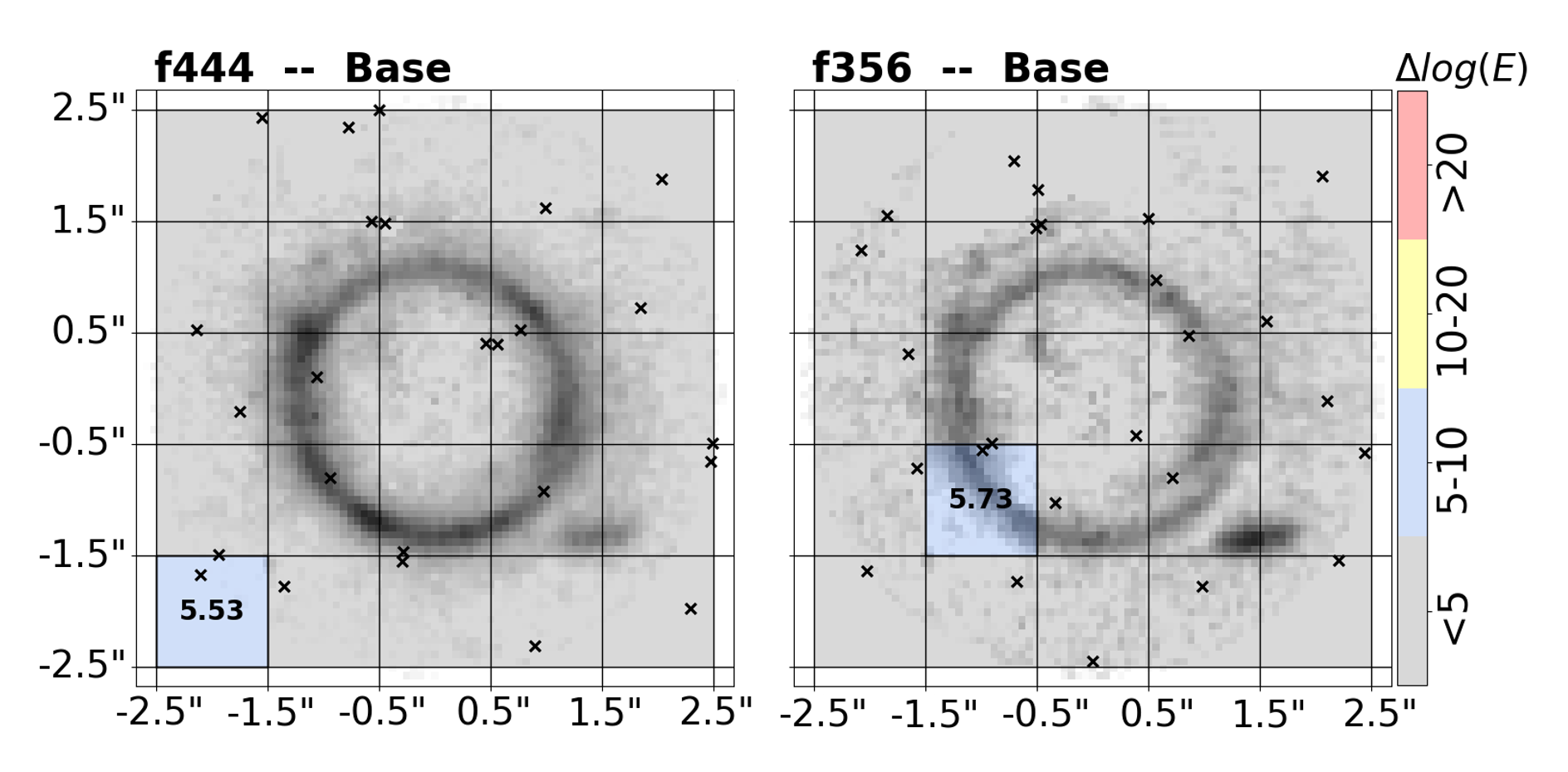}
    \caption{Results from the grid-based preliminary subhalo searches in SPT0418 for the base EPL+Shear model. The values and colours correspond to the Log-Evidence change - \dlev - when a subhalo is modelled in a certain grid cell. Crosses mark the best fit position of the subhalo within each grid cell. The highest evidence grid-cells are marked in bold. \dlev $= 10$ is considered a roughly 5$\sigma$ detection.}
    \label{fig:0418grid_base}
\end{figure*}

Figure \ref{fig:0418grid_base} shows the subhalo grid-search results for the f444 and f356 base models in SPT0418. The f277 filter had too low a source signal level to produce reasonable model results, so it is not shown. There is a potential indication of substructure signal towards the lower left of the image, but in only one grid cell in each filter, and moving closer to the einstein ring in the lower-signal f356 filter. The grid-search evidences are \dlev $ \sim 5$, and the finalised search evidences are \dlev $= 4.44, 4.31$ for f444, f356, meaning there is no appreciable evidence for substructure in this lens when using the base EPL+Shear model.

\subsubsection{Results of Multipole Perturbations}\label{0418mp}

\begin{figure}
    \centering
    \includegraphics[width=0.99\linewidth]{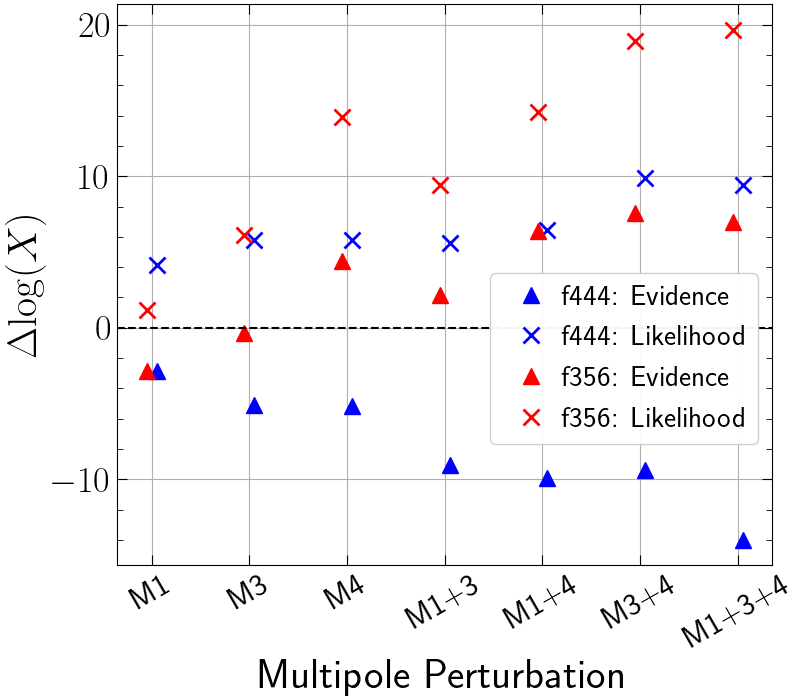}
    \caption{\dlev (Log-Evidence changes, triangles) and $\Delta {\rm log}(\mathcal{L})$ (Log-Likelihood changes, crosses) for the addition of multipole perturbations to the base model of SPT0418.}
    \label{fig:0418_evi_lhd}
\end{figure}

Given that there is no evidence for substructure with the base EPL+Shear model, we already know that the model can fit the data reasonably well without any additional perturbations, and thus we do not expect strong evidences for multipole complexity in the mass model. Figure \ref{fig:0418_evi_lhd} shows \dlev (evidence - triangles) and $\Delta {\rm log}(\mathcal{L})$ (likelihood - crosses) when multipoles are added to the EPL+Shear model. The small positive log-likelihood increases indicate that multipole models produce a small improvement to the overall fit. However, the decrease in evidence implies that this does not outweigh the increase in model complexity and thus the multipole models are disfavoured overall.

The f444 filter has no increase in \dlev for any multipole addition, and in fact, as more multipoles are added, the magnitude of the evidence drop increases - as one would expect if adding further unnecessary complexity to the model. The f356 filter does show minor increases of \dlev $ < 10$ in all additions except for M1-only and M3-only. The fit to the f356 filter however required the model to be initialised from the f444 model as the source signal was too low to provide a reasonable constraint on parameters. We therefore place much less value on the f356 results and show them primarily for completeness, using the f444 as the primary result set. Regardless, the low evidences for multipoles in f356 coupled with the negative evidences for multipoles in f444 lead to the conclusion that there is no basis to claim multipole angular complexity in the mass of lens - an important contrast to SPT2147. We show the parameters for the fits in Appendix \ref{app:additional_params}, Figure \ref{fig:0418_mp_params} and show that they are all consistent across every model and filter, with the exception of f444 shear angles being generally higher than f356. The consistency and the fact that all multipole strengths return close to zero strengthens the opinion of the Base model being most appropriate as the multipole models offer little to no difference.

\subsubsection{Inclusion of both Multipoles and Substructure}\label{0418combi}

\begin{figure}
    \centering
    \includegraphics[width=0.99\linewidth]{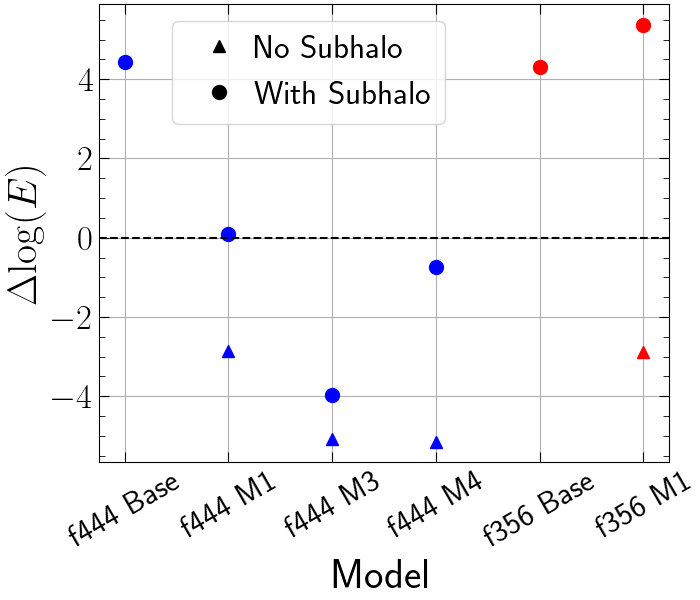}
    \caption{\dlev (Log-Evidence changes) for the addition of a multipole (triangles, for reference), and then subhalos (circles) to the best models for SPT0418. All evidences are compared to the Base no-subhalo smooth model. The dashed line marks \dlev $=0$.}
    \label{fig:0418evi_sub}
\end{figure}

As mentioned, the f356 model fit required initialising from the f444 model results, and therefore we primarily test the f444 model with substructure additions. The f444 model, however, had all negative log-evidence changes upon multipole addition, so we therefore only test the addition of a substructure to the models with a single multipole perturbation. Figure \ref{fig:0418evi_sub} shows the finalised log-evidence changes from the smooth EPL+Shear model for the multipole additions (for reference) and the multipole+subhalo (or just subhalo for the Base model) additions for the f444 Base, M1, M3, M4 models and the f356 Base and M1 models. 

The subhalo grid-searches for the multipole models are not shown as they contain no cells with \dlev $> 10$, and only one cell with \dlev $> 5$ in f444, although interestingly this is now on the lower-right of the lens rather than the lower-left as seen in Figure \ref{fig:0418grid_base}, but is likely just fitting noise or compensating for ther multipole addition. Figure \ref{fig:0418evi_sub} does show that the M1 subhalo fit in f356 has an evidence boost from the smooth M1 model of \dlev $=8.25$, but considering the addition of the M1 causes a drop in evidence of $-2.88$ from the base EPL+Shear model, we suspect the subhalo is primarily just compensating for this.

\subsubsection{Model Residuals}\label{0418_residuals}

\begin{figure*}
    \centering
    \includegraphics[width=0.9\linewidth]{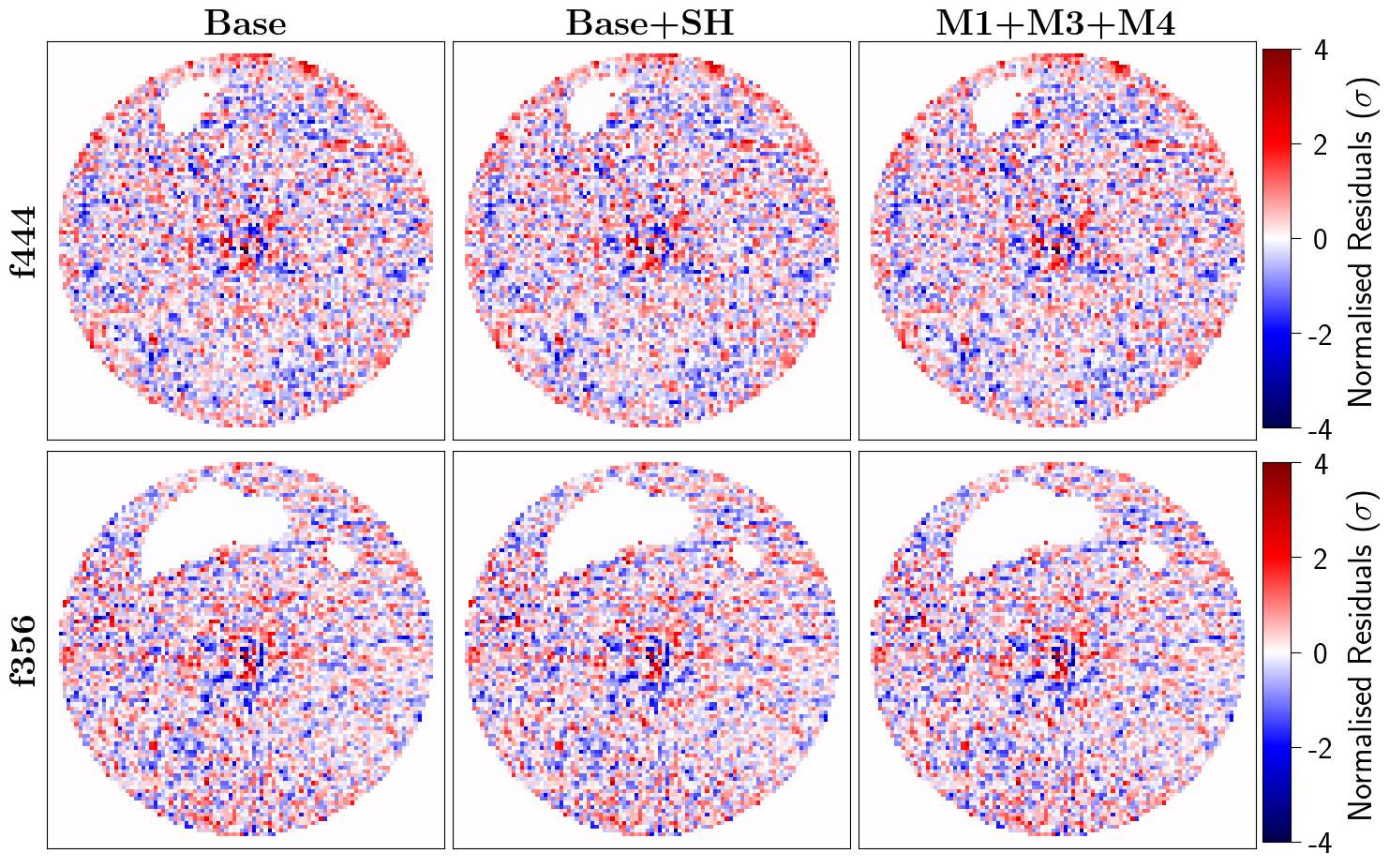}
    \caption{ Normalized residuals (residuals divided by noise in each pixel) from the model fits to SPT0418 for the `Base' (EPL+Shear) model, this model with subhalo (`SH'), and the model containing $m=1$, $m=3$ and $m=4$ multipoles (but no substructure). We see no areas where there are major correlated residual changes when extra complexity (angular or substructure) is added to the model.}
    \label{fig:0418_residuals}
\end{figure*}

We plot the normalized residuals for SPT0418 in Figure \ref{fig:0418_residuals}, showing the Base model, Base + subhalo model, and M1+M3+M4 model. We see that, unlike in SPT2147 (Section \ref{2147_residuals}), there is no noticable change in the residuals when adding either angular complexity or substructure to the Base model, which supports the narrative that this lens does not have detectable complexity.
\section{Discussion}\label{discussion}

\subsection{Is there a dark subhalo in SPT2147?}\label{disc:main}

Our analysis of SPT2147 provides strong evidence (exceeding $10\sigma$) for both: (i) models with angular complexity (multipoles) when a dark matter (DM) subhalo is omitted, and (ii) models including a DM subhalo when multipoles are omitted. The preferred model (see Section \ref{evidence_method}) includes both $m1$ and $m4$ multipoles and a DM subhalo, showing evidence increases of $\Delta \log \mathcal{E} = 10.97$, $9.96$, and $4.3$ for the f444W, f356W and f277W, compared to models without a DM subhalo. This suggests that SPT2147 likely requires both angular complexity in the lens mass distribution and the presence of a DM subhalo. However, systematic uncertainties in strong lens modelling make it difficult to come to a definitive interpretation, leading us to consider three possibilities:

\begin{enumerate}
    \item \textbf{Lens galaxy contains only angular complexity, without a DM subhalo:} In this scenario, the DM subhalo is producing a false positive detection with an evidence increase above $50$ by compensating for missing complexity in the power-law lens mass model, an effect shown in \citet{he2023BPL}. However, the model including $m1$ and $m4$ multipoles still favors the DM subhalo with an evidence increase of $10.97$, and even with all of $m1$, $m3$, and $m4$, the evidence increase for substructure is $9.68$, meaning that if a DM subhalo is truly not present the mass distribution must have complexity beyond what the combination of the $m1$, $m3$ and $m4$ multipoles can capture.
    
    \item \textbf{Lens galaxy contains a DM subhalo, without angular complexity }: Here, the multipole-only models are favored because they are actually fitting the lensing signal of a genuine DM subhalo, a degeneracy outlined by \citet{oriordan2023multipoles}. However, some models with the combination of multipoles and substructure are consistently a higher evidence than the models with substructure only, implying that there is excess lensing potential beyond what can be captured by a single subhalo.
    
    \item \textbf{Lens galaxy contains both angular complexity and a DM subhalo:} A combination of the above cases. The models including a DM subhalo but lacking multipoles see evidence increases of $\sim 50$ because the subhalo primarily accounts for missing complexity in the power-law model of the central mass distribution (see \citealt{he2023BPL}). However, even after the multipoles are added, accounting for this missing complexity, a DM subhalo remains necessary for an improved model fit as there is a true subhalo in the system. 
\end{enumerate}

Option iii above, that this lens contains both substructure and multipoles, is the solution with the highest Bayesian evidence, and is a similar option to recent work by \citet{enzi2024} on SDSSJ0946+1006, who find evidence for significant multipoles alongside the substructure in this lens, but note that the multiple source configuration of this lens is essential to reducing the degeneracy between multipoles and other parameters - a luxury that we do not have with SPT2147-50. 

Given the current evidence, we consider SPT2147 as a dark matter subhalo \textit{candidate}. Future work will aim to refine this interpretation using more flexible lens mass models, such as potential corrections \citep{suyu2009potential, vegetti2012detection}. SPT2147 is an ideal target to investigate this approach, given the apparent need for both angular complexity and a DM subhalo in its modelling, along with the availability of multi-wavelength data for validation. Additionally, follow-up lensing analysis with ALMA data is planned, building on \citet{arisBAR}. Whilst we are not there yet, SPT2147 suggests that separating a dark matter substructure from angular complexity in strongly lensed extended sources is a tractable task, ensuring that it can prove a key test of the dark matter particle in the near future. 

\subsection{Candidate position and mass}\label{disc:pos and mass}

The best-fit position for the dark matter (DM) subhalo in SPT2147 lies approximately 1" north of the uppermost part of the lensed arc. In the f277W filter, the subhalo's position shifts onto the lensed arc, however this is not in tension with the f444W and f356W results, due to greatly increased errors on the position. In the upper panels of Figure \ref{fig:2147subhalo_params} we see that all subhalo x-positions agree to $3\sigma$ except in the M3+4 f277W model, and subhalo y-positions agree except f277W-f444W results with the Base and M1+M3 models - the important point being that there is good agreement to $3\sigma$ in subhalo position (and masses) of our preferred model of M1+M4. The reason for the discrepancy in precision of the f277W results is likely due to the much lower signal-to-noise ratio of the lensed source in this filter (see Table \ref{tab:SNR}) and the resultant need of the model to bring the subhalo closer to the arc to achieve a noticeable perturbation with this low signal. Due to this, we assume that the true subhalo position aligns with the fits from the f444W and f356W bands, at the centre-top of the image, and primarily look at these filters (and specifically f444W) from here forward.

While the subhalo's relatively large inferred mass helps explain how it can still cause a lensing perturbation at this distance from the arc, a DM halo of $\sim 10^{11}M_{\odot}$ would typically be expected to host a visible dwarf galaxy. However, no such galaxy is observed. In Appendix \ref{app:SH_vis}, we estimate the upper limit for the subhalo's brightness using the JWST Exposure Time Calculator \citep{JWST_ETC}, with settings that match those of the observing run \citep{rigby2023templates} for F444W. The result is an expected SNR of approximately 0.16, indicating that the object would be indistinguishable from the background. A dwarf galaxy detection is not unreasonable as an alternative to a dark subhalo; recent work from \citet{ballard2024} showed that the substructure detection in SDSSJ0946+1006 is compatible with a dwarf galaxy which is too faint to be detected given the available data.

The inferred subhalo mass depends significantly on the choice of its density profile. Our analysis assumes a Navarro-Frenk-White (NFW) profile, which has a shallower central density compared to the pseudo-Jaffe profile used in other works \citep[e.g.,][]{vegetti2010detection, vegetti2012detection}. This assumption leads to higher subhalo mass estimates, as seen in the analysis of SDSSJ0946+1006 by \citet{nightingale2023scanning}, where the inferred mass was around $\sim 10^{10.5}M_{\odot}$, compared to masses in the region of $10^{9.5}M_{\odot}$ inferred from studies using more concentrated profiles \citep[e.g.][]{Minor2021}. Additionally, \citet{despali2018LOS} demonstrated that masses derived using NFW profiles map to lower values when assuming pseudo-Jaffe profiles.

Further exploration of the DM subhalo's density profile could suggest a mass closer to $5.0 \times 10^{9}M_{\odot}$ or below if the subhalo is more concentrated. In fact, a recent study from \citet{despali2024} found that for two detections of dark subhalos in the literature, these subhalos prefer a steeper density slope than the NFW profile, so it is likely that our quoted masses are an over-estimate due to their lower concentration. Additionally, if the subhalo lies along the line-of-sight rather than within the lens plane, this could affect both its detectability and the inferred mass \citep[e.g.,][]{he2022LOS}. The availability of multi-wavelength data for SPT2147 makes it an ideal candidate for future investigations into the subhalo's density profile and its potential impact on mass estimates.

\subsection{Substructure detection criteria}\label{disc:criteria}

Previous strong lens studies that assume a power-law plus shear mass model have adopted different Bayesian evidence criteria for identifying dark matter (DM) subhalo candidates. Detection thresholds vary significantly: \citet{vegetti2010detection, vegetti2012detection} use \dlev of 50, \citet{ritondale2019nondetect} use 100, and \citet{nightingale2023scanning} use 10. Similarly to this work, \citet{hezaveh2016detection} claim a detection in a different strong lens with a value of \dlev $=$47.3 for fits using a power-law plus shear plus $m=3$ and $m=4$ multipoles. Many sensitivity mapping studies set thresholds between 50 and 10 \citep{amorisco2021concentration, despali2022sensitivity, ORiordan2023euclid, oriordan2023multipoles}. An increase of 10 corresponds to roughly a $5\sigma$ detection, making this an appropriate benchmark for considering a detection as significant.

Our study shows that the significance of a DM subhalo detection is sensitive to the choice of mass model. For example, in SPT2147, a power-law plus shear model favored a DM subhalo with evidence increases of \dlev$= 60.35, 42.13, 20.54$ in the F444W, F356W, and F277W bands, respectively. In contrast, when including $m=1$ and $m=4$ multipoles, these values dropped to \dlev$= 10.97, 9.96, 4.3$. This pattern is also observed in other analyses. For example, \citet{nightingale2023scanning} found that the evidence for a DM subhalo in SDSSJ0946+1006, a lens with a confirmed substructure, varied from \dlev$=$72.36 for a power-law plus shear model to 22.52 for a decomposed stars-plus-dark matter model. These results suggest that more refined mass models can reduce the evidence threshold needed to detect subhalos, as they better capture the underlying complexity of the lens.

This has important implications for constraints on the mass of the dark matter particle. As mentioned, sensitivity mapping studies often assume a threshold for detection of up to 50, which may be conservative in their estimates of DM subhalo detectability. If improved mass models, and cross-referencing across multipole filters, allow for a lower detection threshold, these forecasts could be adjusted accordingly. This could lead to a scenario where fewer lenses are needed to achieve comparable constraints on the dark matter particle mass, even though more complex mass models are now assumed. A more thorough evaluation of how to establish robust detection criteria is therefore essential for future studies, and whether our implementation, as described in Section \ref{evidence_method}, remains appropriate.

\subsection{Contrasting SPT2147 with SPT0418}\label{disc:0418 contrast}

The contrast in results between SPT2147 and SPT0418 shows that we do not just detect multipoles simply because they are included in the model. SPT0418 is a prime example that actually not all lens mass models require high levels of complexity, and it is important to note that this improves the confidence that the detections in SPT2147 are not systematics where complexity is added to artificially inflate the evidence (i.e the model doesn't just want all the parameters it can get), since that would then also be the case here. Further, as mentioned in Sections \ref{2147_residuals} \& \ref{0418_residuals}, there is a difference in the effects of the complexity additions on the model residuals - where SPT2147 shows evidence of a correlated improvement to the residuals along the lensing arcs, whereas SPT0418 shows no such improvements. This shows the physical meaning behind the changes (or lack thereof) to bayesian evidences that are seen.

Given some of the recovered axis ratio parameters, the SPT0418 lens mass could likely be reasonably well described by a singular isothermal sphere mass model (`SIS': the simplest applicable model which is the EPL with $q=1, \gamma=2$). This dramatic mass simplicity could have numerous explanations: the level of source signal may be too weak to constrain any complexity in the lens mass; the source itself, which is a merger of two elliptical galaxies \citep{cathey2023_0418}, may, despite its large span over the source plane, lack suitably placed features to enable detection of complexity (i.e. reconstruction of a spiral has fewer lensing solutions than reconstruction of a blob); or alternatively, this lensing galaxy may genuinely have little to no angular structure. Regardless, the lack of multipole evidence in SPT0418 makes the indications shown of angular complexity in SPT2147 that much more compelling, and enhances the argument for further study of this lens. What would additionally be interesting is deeper imaging of SPT0418, in order to provide higher levels of source signal in order to re-test the application of multipoles to see if the evidences change.

\section{Summary and Conclusions}\label{conclusions}

We have investigated multi-wavelength images from JWST of two strongly lensed systems to test the effects of angular (multipole) perturbations to the standard elliptical power-law mass model, and the impact of these perturbations on substructure detection. We find that: 1. SPT2147-50 has evidence for angular complexity (of 1st and 4th order multipole combination), with Bayes Factor \dlev $\sim 47, 35, 16$ in f444, f356, f277 imaging respectively; 2. The inclusion of these multipoles reduces the Bayes Factor for substructure by $\sim 46, 16, 18$ for f444, f356, f277; 3. Irrespective of complexity, our highest SNR filter (f444) always gives a substructure detection above $4\sigma$. We will refer to this as a dark matter {\it candidate}, as we still require the non-parametric potential corrections \citep{vegetti2009potential} to definitively rule it out as a mass modelling systematic error. We do not, however, find any evidence for complexity in SPT0418 (angular or substructure), which indicates that the detections in SPT2147 are less likely to be systematics related to the modelling process.

A quantitative summary (see also Table \ref{tab:2147MP} in the appendix) for SPT2147 is:
\begin{itemize}
    \item When assuming only a standard power law plus external shear model for the lens galaxy mass, we favour a dark matter subhalo inclusion with Bayes Factor \dlev  $=56.90, 25.70, 19.84$ in filters f444, f356, f277. These are all well above $5 \sigma$.
    \item We find similarly strong evidence for angular mass complexity in the lens galaxy mass distribution. Inclusion of a combination of first, third and fourth order multipoles is favoured by bayes factors \dlev $=45.19, 33.54, 12.05$ and inclusion of only first and fourth order multipoles is favoured by Bayes Factors \dlev $=46.99, 35.04, 16.39$ (in filters f444, f356, f277 respectively; again all well above $5\sigma$).
    \item Based on the higher source-SNR filters f444 and f356 only, depending on the combinations of multipoles and substructure used, the first order multipole has magnitude in the range $2.43-9.52 \%$; the third order multipole has magnitude $0.29-1.34 \%$; and the fourth order multipole has magnitude $0.30-1.48$.
    \item A dark substructure is still favoured in models that include multipoles, but to a lesser extent. By implementing a `confidence region' of \dlev $< 5$ from the highest evidence model in each filter, the best applicable model (within this region in all three filters) includes only first and fourth order multipoles (plus substructure), and the Bayes Factors for substructure additions are \dlev $=10.97, 9.96, 4.35$ in f444, f356, f277 (approx $5.05, 4.85, 3.4 \sigma$). 
    \item This best model substructure has mass ${\rm log}_{10}M_{200} = 10.87^{+0.53}_{-0.71} M_{\odot}$ in f444 and ${\rm log}_{10}M_{200} = 10.99^{+0.33}_{-0.37} M_{\odot}$ in f356. This is a substructure {\it candidate}, requiring more detailed future analysis to definitively confirm it is not a systematic resulting from an overly simplistic mass model.
\end{itemize}

Our research contributes to the expanding body of evidence that detecting dark matter subhalos through strong lensing necessitates a meticulous separation of the dark matter subhalo signal from the radial and angular complexities inherent in the lens galaxy's mass distribution. This conclusion aligns with previous findings from cosmological simulations \citep{he2023BPL}, observational data \citep{nightingale2023scanning, barone2024lens}, and machine learning applied to large samples of simulated lenses \citep{oriordan2023multipoles}, as well as studies explaining flux ratio anomalies in lensed quasars through mass model angular complexity \citep{gilman_FRA}. Despite fitting one of the most complex parametric mass models currently possible - incorporating a power-law profile with external shear and first, third, and fourth order multipoles - we still detect statistical evidence at a significance level exceeding 5$\sigma$ in favor of the presence of a dark matter subhalo. Although we cautiously refer to this as a \textit{candidate} and acknowledge the need for more advanced lens modeling including potential corrections to reach a definitive conclusion, our findings suggest that disentangling dark matter substructure from angular complexity is an achievable challenge. This progress will eventually enable us to critically assess the cold dark matter paradigm.

\section*{Acknowledgements}

This work was supported by STFC consolidated grants ST/X001075/1 and ST/T000244/1. SL is supported by STFC studentship grant S0600186. CSF acknowledges support by the European Research Council (ERC) through  Advanced Investigator grant to C.S. Frenk, DMIDAS (GA 786910). This work used the DiRAC@Durham facility managed by the Institute for Computational Cosmology on behalf of the STFC DiRAC HPC Facility (www.dirac.ac.uk). The equipment was funded by BEIS capital funding via STFC capital grants ST/K00042X/1, ST/P002293/1, ST/R002371/1 and ST/S002502/1, Durham University and STFC operations grant ST/R000832/1. DiRAC is part of the National e-Infrastructure. This work was also supported using resources provided by the Cambridge Service for Data Driven Discovery (CSD3) operated by the University of Cambridge Research Computing Service, provided by Dell EMC and Intel using Tier-2 funding from the Engineering and Physical Sciences Research Council (capital grant EP/T022159/1), and DiRAC STFC funding.
We would also like to thank the reviewer for their insightful comments which helped improve the paper.

\section*{Software Citations}

This work uses the following software packages:

\begin{itemize}

\item
\href{https://github.com/astropy/astropy}{\texttt{Astropy}}
\citep{astropy1, astropy2}

\item 
\href{https://github.com/ACCarnall/bagpipes}{\texttt{BAGPIPES}}
\citep{bagpipes1}

\item
\href{https://bitbucket.org/bdiemer/colossus/src/master/}{\texttt{Colossus}}
\citep{colossus}

\item
\href{https://github.com/dfm/corner.py}{\texttt{corner.py}}
\citep{corner}

\item
\href{https://github.com/matplotlib/matplotlib}{\texttt{matplotlib}}
\citep{matplotlib}

\item 
\href{https://github.com/johannesulf/nautilus}{\texttt{Nautilus}}
\citep{nautilus}

\item
\href{numba` https://github.com/numba/numba}{\texttt{numba}}
\citep{numba}

\item
\href{https://github.com/numpy/numpy}{\texttt{NumPy}}
\citep{numpy}

\item
\href{https://github.com/rhayes777/PyAutoFit}{\texttt{PyAutoFit}}
\citep{pyautofit}

\item
\href{https://github.com/Jammy2211/PyAutoGalaxy}{\texttt{PyAutoGalaxy}}
\citep{Nightingale2018, pyautogalaxy}

\item
\href{https://github.com/Jammy2211/PyAutoLens}{\texttt{PyAutoLens}}
\citep{Nightingale2015, Nightingale2018, pyautolens}

\item
\href{https://github.com/ljvmiranda921/pyswarms}{\texttt{PySwarms}}
\citep{pyswarms}

\item
\href{https://www.python.org/}{\texttt{Python}}
\citep{python}

\item
\href{https://github.com/scikit-image/scikit-image}{\texttt{scikit-image}}
\citep{scikit-image}

\item
\href{https://github.com/scikit-learn/scikit-learn}{\texttt{scikit-learn}}
\citep{scikit-learn}

\item
\href{https://github.com/scipy/scipy}{\texttt{Scipy}}
\citep{scipy}

\end{itemize}


\section*{Data Availability}

Data for the observations used in this work are publicly available at the STScI MAST database - \url{mast.stsci.edu}, with an overview of the observing project ({\it TEMPLATES}) available at \citep[][doi: \url{10.3847/1538-4357/ad7501}]{rigby2023templates}. The software used for lens modelling, {\tt PyAutoLens}, is open source and available at \url{github.com/Jammy2211/PyAutoLens}. Model parameters and images, including lens and source light subtracted images, are available upon request; please contact the authors.

\bibliographystyle{mnras}
\bibliography{citations}

\newpage
\appendix

\section{Physical interpretation of multipoles}\label{app:mulitpoles}

\begin{figure*}
    \centering
    \includegraphics[width=0.9\linewidth]{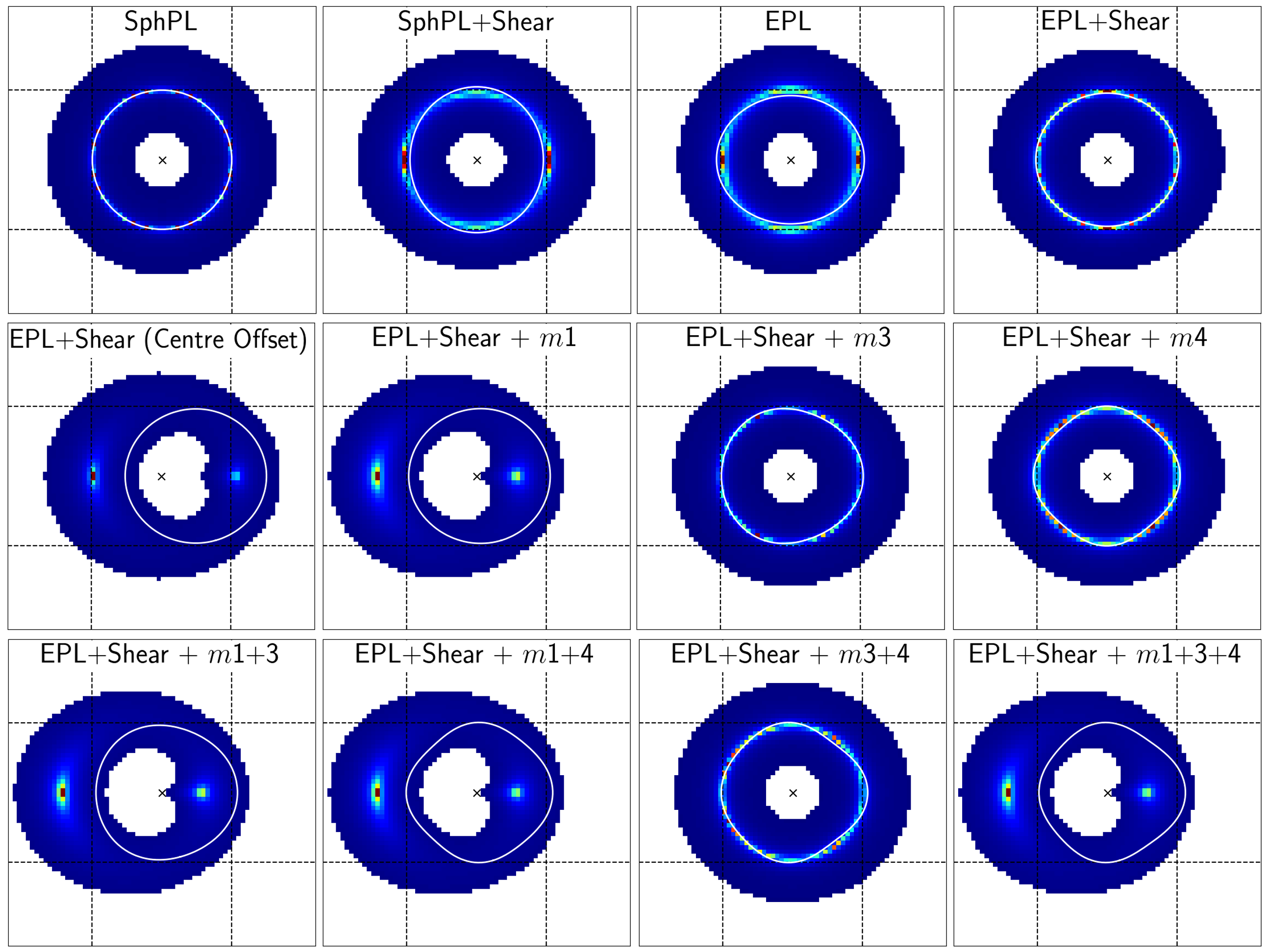}
    \caption{A simple system (spherical sersic source at $z=4$ perfectly aligned with spherical Power-Law mass, slope=2.1, $z=0.5$), showing the effects of changes to centre (${\Delta}x=+0.5 {\rm arcsec}$), mass ellipticity ($q=0.85$), addition of external shear ($k_{\gamma}=0.05$), and addition of multipole perturbations of order $m=1,3,4$ with strengths $k_m=0.05, 0.01, 0.01$. The image is a noiseless image, the white lines represent the lensing Critical Curves, and the dashed lines are drawn for reference at ${\pm}1 \, {\rm arcsec}$, with the origin marked with the black cross. All angles in the simulation are set to zero.}
    \label{fig: multipoles}
\end{figure*}

Figure \ref{fig: multipoles} shows a simple example of the effects on the lensing of a simple Sersic-light source of the first 4 multipole orders when added to a spherical Power-Law mass distribution with ${\rm slope}=2.1$. Note that we demonstrate a multipole strength of 0.1 for $m=1$ and 0.05 for $m=3, m=4$ as the $m=1$ multipole requires a greater strength to produce its effects \citep{arisM1}. Also note that in these perfect systems the multipole perturbation creates $2m$ multiple images, however, if we added noise, lens light and source complexity, some of these multiple images will blur back into the Einstein ring based on the system configuration.

At $\phi = 0$, the $m=2$ multipole gives a similar effect to the Power-Law mass ellipticity and at $\phi = 90^\circ$, it is similar to the External Shear, and could explain therefore the indications in studies such as \citet{etherington2023strong} that the External Shear is fitting to internal lens mass complexity. For a physical interpretation of the remaining multipoles, we visualise that the $m=1$ multipole corresponds to a skew in the mass distribution, which could, for example, be due to the galaxy being mid- or post-merger. The $m=3$ and $m=4$ multipoles are more frequently used in strong lensing \citep[e.g.][]{hezaveh2016detection, stacey2024complex} and demonstrate triaxiality from $m=3$, which has long been a component of galaxy surface brightness fits \citep[e.g.][]{benacchio1980triaxiality, tremblay1995m3}, and boxiness/disciness from the $m=4$ \citep[e.g.][]{bender1989m4}. The majority of physical intuition on these multipoles comes from the light profiles of galaxies - see e.g. \citet{hao2006shapes, padilla2008shapes} for comprehensive applications of multipoles to Sloan Digital Sky Survey galaxies, or \citet{cappellari2016shapes} for a review on galaxy structure using Spectroscopy. Note that work is ongoing to achieve a quantitative rather than qualitative view on what multipole strengths and realisations are to be expected directly from mass modelling of strong lensing galaxies \citep{arisM1, stacey2024complex}, but so far in the literature we see that the amplitudes of mass multipoles are in line with those from the light, but potentially mis-aligned \citep{stacey2024complex}.

\section{Calculation of substructure visibilty}\label{app:SH_vis}

\begin{figure}
    \centering
    \includegraphics[width=1\linewidth]{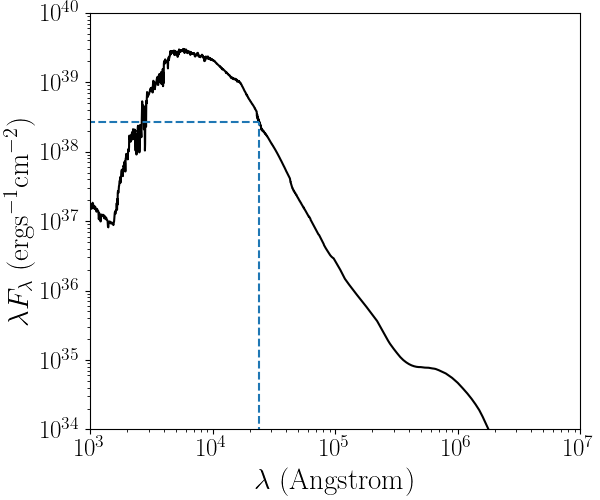}
    \caption{Spectral Energy Distribution (SED) based on a 5-year old Elliptical Galaxy from the SWIRE library. The SED has been scaled using our calculated luminosity to represent the true non-normalised expected values of $\lambda F_{\lambda}$. The blue dashed lines trace the values at our emitted light wavelength, $\lambda_{\rm emit}=2.41 {\mu}m$.}
    \label{fig:SED}
\end{figure}

To approximate if the theorised subhalo is detectable from its stellar emission (in the NIRCam f444W filter, where we have the highest source signal), we first start with the best fit subhalo mass of $M_{200}=10^{10.87}M_{\odot}$. We operate under some simplifying assumptions that there is no dust extinction and that the halo has retained its original mass (in reality dust would cause a lower SNR for the object, but if the halo is at the lens redshift, there would be tidal stripping which reduced its mass over time, meaning it could have a larger stellar component than calculated).  Following the formalism of the Stellar-to-Halo mass relation proposed by \citet{moster2010SHMR} in equation \ref{eq:SHMR}, and taking the best-fit parameters from Table 1 of \citealt{Girelli2020SHMF} at $z=0.845$, we derive a rough estimate of $\log_{10}(M_{*}/M_h){\sim} -2.22$ or ${\rm log}_{10}{M_*}{\sim}8.65M_\odot$. Taking a mass-to-light ratio of $1$, this also gives a luminosity of ${\rm log}_{10}(L_*) \sim 8.65L_\odot$. 

\begin{equation}
    \label{eq:SHMR}
    \frac{M_*}{M_h}(z) = 2A(z) \left[ \left( \frac{M_h}{M_{A(z)}} \right)^{-\beta (z)} + \left( \frac{M_h}{M_{A(z)}} \right)^{\gamma (z)}  \right]^{-1}
\end{equation}

We take an example Spectral Energy Distribution (SED) from the SWIRE\footnote{\url{https://www.iasf-milano.inaf.it/~polletta/templates/swire_templates.html}} template library \citep{polletta2007spectral} for a 5-Gyr Elliptical Galaxy, with the SED shown in figure \ref{fig:SED}. This is only an approximation of the SED, but a reasonable one, particularly regarding age given the lookback time of $\sim7$ Gyr. 
We estimate the Luminosity distance at this redshift using the {\tt CosmoTools}\footnote{\url{http://www.bo.astro.it/~cappi/cosmotools}} cosmological calculator, this gives $D_L=5292 {\rm Mpc}$ for $H_0=71, \Omega_M=0.3, \Omega_{\Lambda}=0.7$. Using this distance, we calculate the rest-frame wavelength for light at this redshift when observed in the centre of the f444 filter is ${\lambda}_{\rm , emit}=2.41 {\mu}m$, and using the SED we get $L_{\lambda {\rm , emit}}=3.65*10^{32} {\rm erg s}^{-1} \textup{~\AA}^{-1}$, which converts to an observed flux density at $\lambda = 4.44\mu m$ of $F_{\lambda {\rm , obs}}=1.089*10^{-23} {\rm erg s}^{-1}{\rm cm}^{-2} \textup{~\AA}^{-1}=$.

We use the JWST Exposure Time Calculator (ETC)\footnote{\url{https://https://jwst.etc.stsci.edu/}} \citep{JWST_ETC} to then calculate the upper-limit signal-to-noise we would expect from an actual observation (we do not apply any extinction). We apply the same observing strategy from the {\it TEMPLATES} \citep{rigby2023templates} observing proposal\footnote{\url{https://www.stsci.edu/jwst/phase2-public/1355.pdf}}, approximate the light as a point source, and use the same built-in SED of a 5-Gyr old elliptical galaxy to get the observed filter flux. We also use the ETC ability to extract and use a noise map generated from the day of the observation (09-Sep-2022) and implement this to output the expected signal-to-noise ratio for this object to be ${\rm SNR} \sim 0.16$.

\section{Additional fit parameter plots}\label{app:additional_params}

\begin{figure*}
    \centering
    \includegraphics[width=0.99\linewidth]{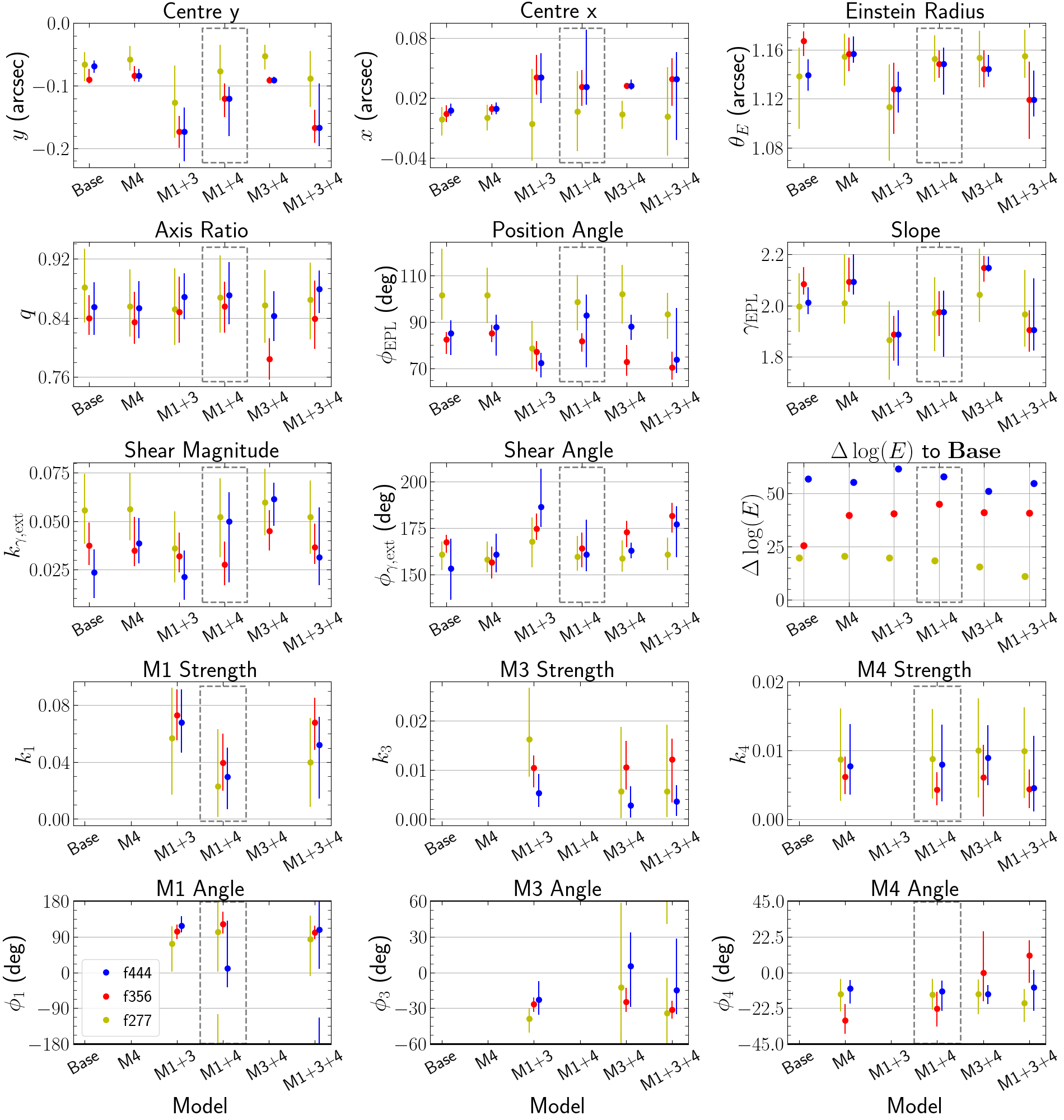}
    \caption{Reported parameters and associated errors for the fits after adding a subhalo to the highest evidence macro-models in SPT2147, as well as the Base model. \dlev shown here is the total difference to the Base no-subhalo model. Multipole angles have a symmetry through $360/n$ - i.e. a 90 degree error on the M4 multipole means it is unconstrained as this multipole has rotational symmetry through 90 degrees, and so the errors are wrapped around $\pm (360/n)/2$. Evidence changes shown are in comparison to the Base EPL+Shear model. We highlight with the gray box our preferred final model of EPL+Shear+M1+M4.}
    \label{fig:2147_sh_params}
\end{figure*}

We show here first the parameters from the finalised fits in each filter for the subhalo models of SPT2147 in figure \ref{fig:2147_sh_params}, with the prime M1+M4 model highlighted. Most parameters are consistent between filters within a model, but there are some $3\sigma$ tensions, including importantly f277-f356 position angle tension in the primary M1+M4 model. Additional tensions beyond $3\sigma$ are: f356-f444 tension in the Einstein radius for the Base model; f277-f444 tension in the position angle for the Base model, and f277-f356 position angle tension in M4, (M1+M4), M3+M4 and M1+M3+M4 models; f277-f356 tension in the shear angle for the M1+M3+M4 model.  

\begin{table*}
    \centering
    \begin{tabular}{cccccccccccc}
         & & \textbf{PRE-sub} & \textbf{PRE-sub} & \multicolumn{3}{c}{\textbf{PRE-sub}} & \textbf{WITH-sub} & \textbf{WITH-sub} & \multicolumn{3}{c}{\textbf{WITH-sub}} \\
         
        \textbf{Filter} & \textbf{Model} & \textbf{\dlev} & \textbf{Max Normalised} & \multicolumn{3}{c}{\textbf{Multipole Strengths}} & \textbf{\dlev} & \textbf{Max Normalised} & \multicolumn{3}{c}{\textbf{Multipole Strengths}} \\
        
         &  & (to Base) & \textbf{Residual ($\sigma$)} & \multicolumn{3}{c}{\textbf{($k_m *10^{-2}$)}}  & (to Previous) & \textbf{Residual ($\sigma$)} & \multicolumn{3}{c}{\textbf{($k_m *10^{-2}$)}} \\
         
         &  &  &  & \textbf{m1} & \textbf{m3} & \textbf{m4} &  &  & \textbf{m1} & \textbf{m3} & \textbf{m4} \\
        \hline
         f444W & Base   & -     & 3.58 & -    & -    & -    & 56.90 & 3.59 & -    & -    & -    \\
         f444W & M1     & 2.36  & 3.67 & 4.98 & -    & -    & 55.83 & 3.71 & 3.94 & -    & -    \\
         f444W & M3     & 21.81 & 3.63 & -    & 0.94 & -    & 28.98 & 3.60 & -    & 0.29 & -    \\
         f444W & M4     & 43.13 & 3.66 & -    & -    & 1.48 & 12.15 & 3.59 & -    & -    & 1.11 \\
         f444W & M1+3   & 48.16 & 3.73 & 9.52 & 0.94 & -    & 13.36 & 3.74 & 6.79 & 0.54 & -    \\
         f444W & M1+4   & 46.99 & 3.59 & 2.81 & -    & 1.41 & 10.97 & 3.50 & 2.96 & -    & 0.80 \\
         f444W & M3+4   & 38.01 & 3.58 & -    & 0.57 & 0.98 & 13.03 & 3.54 & -    & 0.29 & 0.90 \\
         f444W & M1+3+4 & 45.19 & 3.59 & 2.43 & 0.49 & 1.10 & 9.68  & 3.75 & 5.23 & 0.36 & 0.46 \\
         \hline
         f356W & Base   & -     & 3.83 & -    & -    & -    & 25.70 & 3.77 & -    & -    & -    \\
         f356W & M1     & 10.66 & 3.94 & 5.99 & -    & -    & 28.77 & 4.05 & 4.87 & -    & -    \\
         f356W & M3     & 4.79  & 3.80 & -    & 0.67 & -    & 40.99 & 3.87 & -    & 1.34 & -    \\
         f356W & M4     & 24.31 & 3.88 & -    & -    & 0.92 & 15.57 & 3.76 & -    & -    & 0.62 \\
         f356W & M1+3   & 38.69 & 4.04 & 8.11 & 0.84 & -    & 2.04  & 3.99 & 7.30 & 1.04 & -    \\
         f356W & M1+4   & 35.04 & 4.08 & 5.97 & -    & 0.76 & 9.96  & 4.06 & 3.97 & -    & 0.43 \\
         f356W & M3+4   & 18.61 & 3.85 & -    & 0.31 & 0.96 & 22.40 & 3.87 & -    & 1.05 & 0.61 \\
         f356W & M1+3+4 & 33.54 & 4.08 & 7.62 & 0.62 & 0.30 & 7.31  & 4.03 & 6.78 & 1.21 & 0.44 \\
         \hline
         f277W & Base   & -     & 3.47 & -    & -    & -    & 19.84 & 3.49 & -    & -    & -     \\
         f277W & M1     & 4.10  & 3.48 & 5.30 & -    & -    & 13.82 & 3.49 & 3.51 & -    & -     \\
         f277W & M3     & 5.68  & 3.44 & -    & 1.72 & -    & 10.58 & 3.47 & -    & 1.37 & -     \\
         f277W & M4     & 16.29 & 3.72 & -    & -    & 1.18 & 4.25  & 3.46 & -    & -    & 0.87  \\
         f277W & M1+3   & 12.99 & 3.45 & 9.00 & 1.33 & -    & 6.78  & 3.48 & 5.67 & 1.63 & -     \\
         f277W & M1+4   & 16.39 & 3.56 & 3.58 & -    & 1.20 & 2.07  & 3.47 & 2.31 & -    & 0.88  \\
         f277W & M3+4   & 11.23 & 3.51 & -    & 0.46 & 1.10 & 4.35  & 3.72 & -    & 0.57 & 1.00  \\
         f277W & M1+3+4 & 12.05 & 3.55 & 4.12 & 0.54 & 1.00 & -0.99 & 3.49 & 3.99 & 0.57 & 0.99  \\
         \hline
    \end{tabular}
    \caption{Evidence boosts, maximum normalised residuals, and multipole strengths for SPT2147 over all models both with and without the addition of a Subhalo. All multipole 'pre-sub' evidences are compared to the Base 'pre-sub' evidence. All 'post-sub' evidences are compared to the corresponding model without a subhalo.}
    \label{tab:2147MP}
\end{table*}

\begin{figure*}
    \centering
    \includegraphics[width=0.99\linewidth]{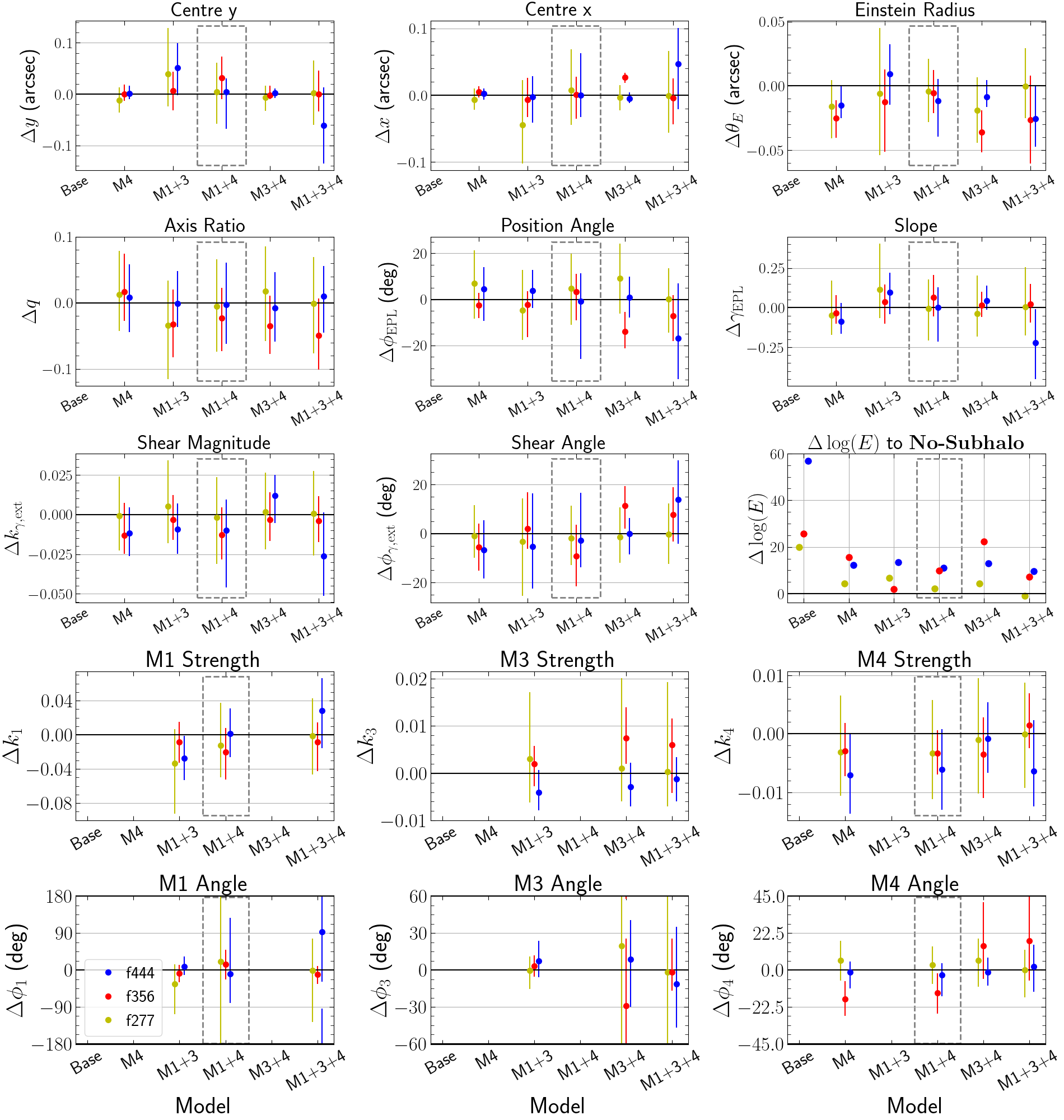}
    \caption{Changes in the mass macro-model parameters for SPT2147 once we add a substructure, as well as the associated error on the change. Solid lines are drawn at a change of $0$. Multipole angles have a symmetry through $360/n$ - i.e. a 90 degree error on the M4 multipole means it is unconstrained as this multipole has rotational symmetry through 90 degrees, and so the errors are wrapped around $\pm (360/n)/2$. Evidence changes shown are the total difference to the Base no-subhalo model. We highlight with the gray box our preferred final model of EPL+Shear+M1+M4.}
    \label{fig:2147_sh_changes}
\end{figure*}

Next, in Figure \ref{fig:2147_sh_changes} we show the change in the parameters of the fits from before adding a subhalo to after this addition, with the prime M1+M4 model highlighted. We draw solid lines at 0 to show the lines of no change - errors are the quadrature sum of the parameter $3\sigma$ errors from the fits before and after and if errorbars cross the zero-line, the parameter change is consistent with 0. There are some non-zero-consistent parameters, but all M1+M4 prime model parameters are consistent before and after the subhalo addition, excepting a small negative change in the f356 model for the M4 angle. Parameters with changes outside of zero are: f444 Base and f356 M3+M4 positive change of centre x position; f277, f356, f444 Base, f356 M4, f356 M3+M4, f444 M1+M3+M4 decrease in einstein radius; f277 Base increase and f356 M3+M4 decrease in position angle; f444 Base and f444 M1+M3+M4 decrease in power-law slope; f444 Base decrease in shear magnitude; f356 M3+M4 increase in shear angle; f444 M1+M3 decrease in M1 strength, f356 M3+M4 increase in M3 strength; f356 M4 and f356 M1+M4 decrease in M4 angle.

We also show in Table \ref{tab:2147MP} specific values for the \dlev evidence changes, maximal normalised residuals from the fits, and multipole strengths, for all the models fit both with and without the subhalo.

\begin{figure*}
    \centering
    \includegraphics[width=0.99\linewidth]{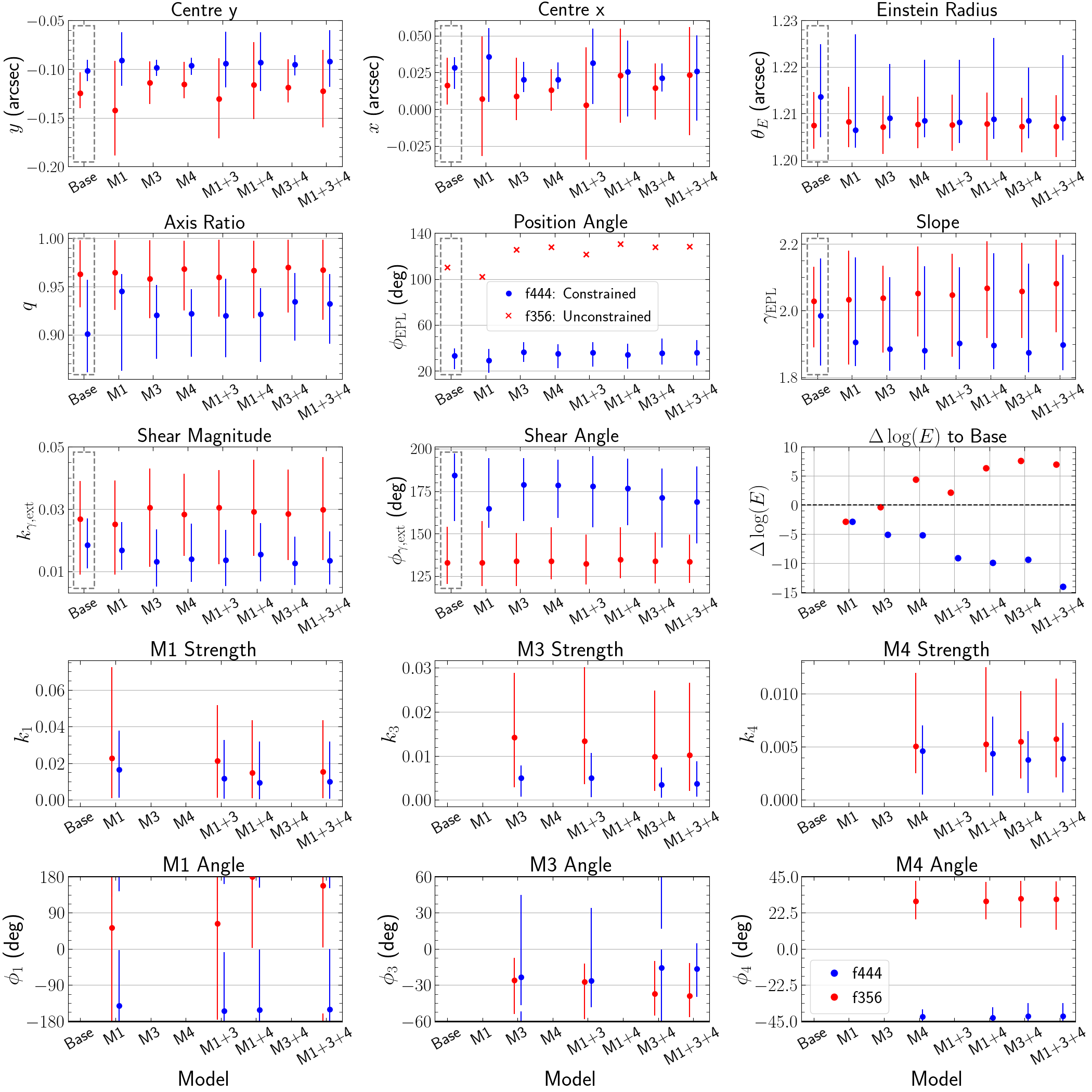}
    \caption{Reported parameters and associated errors for all fits (without a subhalo) to SPT0418. Multipole angles have a symmetry through $360/n$ - i.e. a 90 degree error on the M4 multipole means it is unconstrained as this multipole has rotational symmetry through 90 degrees, and so the errors are wrapped around $\pm (360/n)/2$. Evidence changes shown are in comparison to the Base EPL+Shear model. We highlight with the gray box our preferred final model (Base). The position angles in f356 show only the median value as the errors span the full parameter space leaving this angle unconstrained - this is due to the highly spherical axis ratio.}
    \label{fig:0418_mp_params}
\end{figure*}

We finally show in figure \ref{fig:0418_mp_params} the parameters for the fits to SPT0418. We recall that no multipole models provide a high enough evidence increase to be worth using, so the prime model here is the Base model. The position angles for the f356 fits are unconstrained (errors span the whole range), due to the high axis ratio - $q=1$ is spherical so the position angle has no physical meaning at this point. Shear angles are generally inconsistent between the f356 and f444 filters within a model, with f444 being higher. All other parameters are fully consistent between all models and filters, which is expected when all models have a similar Bayesian evidence. All multipole strengths reach down very close to zero with their errorbars, again showing that the multipole models are not required for this lens.

\bsp	
\label{lastpage}
\end{document}